%% file: main.tex
\documentclass[letterpaper,12pt]{article} 

\usepackage{amsmath}
\usepackage{amssymb}
\usepackage[comma,longnamesfirst]{natbib}
\usepackage[dvips]{epsfig}
\usepackage{dcolumn}
\usepackage{enumerate}
\usepackage{hhline}
\usepackage{dsfont}
\usepackage{afterpage}
\usepackage{arydshln}
\usepackage{graphicx}
\usepackage{color}
\usepackage[usenames,dvipsnames]{xcolor}
\usepackage{rotating}
\usepackage[breaklinks]{hyperref}
\usepackage{breakurl} 
\usepackage{xr}
\usepackage[percent]{overpic}
\usepackage{subfig}
\usepackage[font=footnotesize]{caption}
\usepackage{comment}
\includecomment{commenta} 
\excludecomment{commentb}  

\usepackage{verbatim}
\usepackage{algorithmicx}
\usepackage[noend]{algpseudocode}
\usepackage{algorithm}
\usepackage{diagbox}
\usepackage{graphicx}
\usepackage{wrapfig}
\usepackage{lscape}
\input epsf
\usepackage{fontenc}
\usepackage{setspace}
\usepackage{bm}
\usepackage{slashbox}
\usepackage{lscape}
\usepackage{breakurl} 
\usepackage{multirow}
\usepackage{eurosym}
\usepackage{titlesec}
\epsfverbosetrue
\input{defs_letter}

\input{defs.tex}

\input{alm.tex}

\usepackage[margin=1in]{geometry}
\titlespacing*\section{0pt}{0pt plus 4pt minus 2pt}{0pt plus 2pt minus 2pt}
\titlespacing*\subsection{0pt}{0pt plus 4pt minus 2pt}{0pt plus 2pt minus 2pt}
\titlespacing*\subsubsection{0pt}{0pt plus 4pt minus 2pt}{0pt plus 2pt minus 2pt}

\newcounter{myremark}

\newcounter{mynotation}

\usepackage{paralist}

\renewenvironment{enumerate}[1]{\begin{compactenum}#1}{\end{compactenum}}

\makeatletter
\def\@seccntformat#1{\@ifundefined{#1@cntformat}%
	{\csname the#1\endcsname\quad}  
	{\csname #1@cntformat\endcsname}
}
\let\oldappendix\appendix 
\renewcommand\appendix{%
	\oldappendix
	\newcommand{\section@cntformat}{\appendixname~\thesection\quad}
}
\makeatother

\usepackage{titlesec}

\usepackage{scalerel,stackengine}
\stackMath
\newcommand\reallywidehat[1]{%
\savestack{\tmpbox}{\stretchto{%
  \scaleto{%
    \scalerel*[\widthof{\ensuremath{#1}}]{\kern-.6pt\bigwedge\kern-.6pt}%
    {\rule[-\textheight/2]{1ex}{\textheight}}
  }{\textheight}%
}{0.5ex}}%
\stackon[1pt]{#1}{\tmpbox}%
}

\begin{document}
\setlength{\abovedisplayskip}{0.15cm}
\setlength{\belowdisplayskip}{0.15cm}
\pagestyle{empty}
\input{title}

\newpage
\pagestyle{plain}
\setcounter{equation}{0}
\renewcommand{\theequation}{\arabic{equation}}
\input{sec1}

\input{sec2}

\input{sec3}

\input{sec4}
\input{sec5}

\input{sec6}
\newpage
\input{append}

\newpage
\singlespacing
\bibliography{references}
\newpage

\input{tabs}
\newpage

\input{figs}
\end{document}

%% file: defs_letter.tex
\def\theequation{\thesection.\arabic{equation}}  
\def\abstract{\if@twocolumn
\section*{Abstract}
\else \normalsize 
\begin{center}
{\bf Summary\vspace{-.5em}\vspace{0pt}} 
\end{center}
\quotation 
\fi}
\def\endabstract{\if@twocolumn\else\endquotation\fi}

\makeatletter
\newcommand{\myappendix}[1]{
	\setcounter{section}{1}
        \renewcommand{\thesection}{A\arabic{section}}}

%% file: defs.tex
\def \dsR {\text{$\mathds{R}$}}

\DeclareMathOperator*{\argmin}{{arg\,min}}

\def \xvec {\text{\boldmath$x$}}    
    \def \mY {\text{\boldmath$Y$}}

\def \etavec          {\text{\boldmath$\eta$}}

\def \varthetavec     {\text{\boldmath$\vartheta$}}

%% file: alm.tex
\usepackage{color}
\usepackage{colordvi}
\newlength{\breite}
\breite\textwidth
\newcounter{aufg}[section]
  {\refstepcounter{aufg}\noindent\textbf{Exercise \arabic{aufg}:}
   \\*[1ex]\noindent}{\vspace{.5cm}}
   
 \newcounter{notes}[section]
  {\refstepcounter{aufg}\noindent\textbf{}
   \\*[1ex]\noindent}{\vspace{.5cm}}
   
\usepackage{amsthm}  

\theoremstyle{definition}

\newtheorem*{beisp*}{Example}
\newtheorem{Proof}{Proof}


\newtheoremstyle{break}
  {}
  {}
  {}
  {}
  {\bfseries}
  {.}
  {\newline}
  {}
  
\theoremstyle{break}

\newcommand{\head}[2]%
 {\hrule \vspace{.15cm} {\sfbold Advanced Statistical Inference, Summer Term 2012, Georg-August-University G\"ottingen}\hfill
{\sfbold Sheet #1}\\
{\sfbold Prof. Dr. Thomas Kneib, Nadja Klein}\hfill {\sfbold #2}

\vspace{.2cm}
\hrule

\vspace{1cm}

}

\newcounter{auf}
{\refstepcounter{auf}
\begin{center}
\fcolorbox[gray]{0}{.95}{
\makebox[\breite]{
\textbf{Exercise \arabic{auf}}
}}\\*[1ex]\noindent
\end{center}
}{\vspace{.5cm}}

\newcounter{loes}[section]
{\stepcounter{loes}
\begin{center}
\fcolorbox[gray]{0}{.95}{
\makebox[\breite]{
\textbf{L"osung \arabic{loes}}
}}\\*[1ex]\noindent
\end{center}
}{}

{\begin{center}
\fcolorbox[gray]{0}{.95}{
\makebox[\breite]{
\textbf{Zu Aufgabe #1}
}}\\*[1ex]\noindent
\end{center}\vspace{1cm}
}{\vspace{1cm}}

\newcounter{ka}
{\refstepcounter{ka}
\begin{center}
\framebox[\textwidth]{
\textbf{Aufgabe \arabic{ka}} \hfill #1 Punkte
}\\*[1ex]\noindent
\end{center}
}{\vspace{1cm}}

\newcounter{lka}
{\refstepcounter{lka}
\begin{center}
\framebox[\textwidth]{
\textbf{L\"osung \arabic{lka}} \hfill #1 Punkte
}\\*[1ex]\noindent
\end{center}
}{\vspace{1cm}}


%% file: title.tex
\begin{titlepage}

\title{Boosting Distributional Copula Regression}

\author{Nicolai Hans, Nadja Klein, Florian Faschingbauer, Michael Schneider \\ and Andreas Mayr}

\date{}
\maketitle
\noindent
{\small Nicolai Hans is PhD student and Nadja Klein is Emmy Noether Research Group Leader in Statistics and Data Science at Humboldt-Universit\"at zu Berlin. Andreas Mayr is Professor for Epidemiology and Vice Director of the Department of Biometry, Informatics and Epidemiology at the University Hospital Bonn. Correspondence should be directed to~Prof.~Dr.~Nadja Klein at Humboldt Universit\"at zu Berlin,
Unter den Linden 6, 10099 Berlin. Email: nadja.klein@hu-berlin.de.\\

\noindent \textbf{Acknowledgments:} Nicolai Hans, Nadja Klein and Andreas Mayr gratefully acknowledge support  by the German research foundation (DFG) through the grants KL3037/2-1, MA7304/1-1 (428239776).}\\

\newpage
\begin{center}
\mbox{}\vspace{2cm}\\
{\LARGE \title{Boosting Distributional Copula Regression}
}\\
\vspace{1cm}
{\Large Abstract}
\end{center}
\vspace{-1pt}
\onehalfspacing
\noindent
Capturing complex dependence structures between outcome variables (e.g., study endpoints) is of high relevance in contemporary biomedical data problems and medical research. Distributional copula regression provides a flexible tool to model the joint distribution of multiple outcome variables by disentangling the marginal response distributions and their dependence structure. In a regression setup each parameter of the copula model, i.e.~the marginal distribution parameters and the copula dependence parameters, can be related to covariates via structured additive predictors. We propose a framework to fit distributional copula regression models via a model-based boosting algorithm. Model-based boosting is a modern estimation technique that incorporates useful features like an intrinsic variable selection mechanism, parameter shrinkage and the capability to fit regression models in high dimensional data setting, i.e.~situations with more covariates than observations.
Thus, model-based boosting does not only complement existing Bayesian and maximum-likelihood based estimation frameworks for this model class but rather enables unique intrinsic mechanisms that can be helpful in many applied problems.
The performance of our boosting algorithm in the context of copula regression models with continuous margins is evaluated in simulation studies that cover low- and high-dimensional data settings and situations with and without dependence between the responses. Moreover, distributional copula boosting is used to jointly analyze and predict the length and the weight of newborns conditional on sonographic measurements of the fetus before delivery together with other clinical variables.

\vspace{20pt}

\noindent
{\bf Keywords}: Archimedian copula; compententwise gradient boosting; early stopping; GAMLSS; tail dependence.
\end{titlepage}

%% file: sec1.tex

\section{Introduction}\label{sec:intro}

The analysis of complex association structures between multiple outcome variables is of increasing interest in contemporary biomedical research.
For instance, in genetic epidemiology the joint consideration of multiple phenotypes leads to a better understanding of physical and mental disorders and the identification of relevant genetic risk factors \citep[][]{OttWan2011, SuoTouElv2013, Gho2014}. 
Moreover, in clinical medicine multivariate investigations provide a broader view on diseases like diabetes \citep[][]{EspCadKne2019} or Alzheimer's disease \citep[][]{YanLiWan2015} and important clinical measures like the body fat percentage \citep[][]{PetLauDaS2021}.
The motivating example of this paper is to model and predict the height and weight of newborn babies based on  sonographic and clinical covariates of fetuses and mothers collected at the Erlangen University Hospital before birth \citep{FasDamRaa2016, FasRaaHei2016}. 
In clinical obstetrics and gynecology the prediction of the fetal weight is of high relevance for decision making during the birth process. 
We investigate the fetal weight and height together, as these measures are likely to interrelate. 
From a delivery management perspective the analysis of both responses yields new insights like the identification of cases with disproportional growth or the calculation of joint probabilities that crucial thresholds are passed. Moreover, it could also be of particular interest to explore predictor variables that influence the association between the weight and length of the fetus.\\
To define an appropriate and flexible joint model we employ copulas \citep{Nel2006} and integrate their use into a bivariate distributional regression model. Copula models offer an extremely flexible approach to define multivariate distributions of several outcome variables. 
In particular, by using copulas the model-building process is conveniently decomposed into the specification of the univariate marginal distributions (i.e.~in our case those of fetal weight and height) and the selection of an appropriate copula function that defines the dependence structure \citep[][]{Joe1997,Nel2006}.
Thus, response variables with potentially different marginal distributions can be combined via copula functions that introduce distinct dependence scenarios.
This makes the copula approach a versatile tool for multivariate analysis going well beyond Gaussian distribution assumptions of the marginals or linear correlations. 
Within the structured additive distributional regression framework like generalized additive models for location scale and shape (GAMLSS) \citep{RigSta2005}, different types of univariate response variables, i.e.~continuous, discrete or mixed continuous-discrete, can be considered.
This model class extends the generalized additive model (GAM) \citep{HasTib1990} by associating every response distribution parameter with the covariates via additive predictors. 
The additive predictors allow for different covariate effects, i.e.~linear, non-linear, random and spatial, on the model parameters \citep{Woo2017}.
While there exists a rich literature regarding copula regression \citep[][]{KolPai2009, CraSab2012, KraBreSil2013, SabWeiMia2014, RadMarWoj2016}, these typically cover only parts of the flexibility of the distributional copula regression frameworks introduced in a Bayesian \citep{KleKne2016} and a penalized-likelihood \citep{MarRad2017} based context.
As part of distributional copula regression all parameters of the model, i.e.~both the distribution parameters of the marginals and the dependence parameters, are related to the covariates via additive predictors, which are estimated simultaneously. 
Hence, also the dependence structure is described by flexible covariate effects that possibly exceed classical linear relations.
Note that there exists also a two-step estimation technique, i.e.~first the margins and then the copula, for distributional copula regression models \citep{VatCha2015, VatNag2018}.\\
We are the first contribution in the literature postulating a general boosting approach to structured additive distributional copula regression. We refer to this framework as \textit{boosted copula regression} and we conceptualize and implement a first scenario with continuous marginals and three different copula functions. To do so, we extend a popular estimation technique of modern data analysis called model-based boosting to fit distributional copula regression models.
While the original concept of boosting arose in the machine learning community \citep{FreSch1996}, \citet{FriHasTib2000} provide a statistical view on boosting. 
Model-based boosting \citep{BuhYu2003,BuhHot2007} builds on a functional gradient boosting paradigm \citep[][]{Fri2001} and serves well for estimation problems in the context of univariate regression models. \citet{MayFenHof2012} further extend the model-based boosting algorithm to the GAMLSS class, i.e.~to distributional regression situations. 
The estimation via model-based boosting yields several advantages compared to the Bayesian and frequentist counterparts that are of high interest in applied data analysis. 
First, model-based boosting incorporates an intrinsic mechanism for variable selection. Variable selection, i.e.~the problem to select a small set of truly informative covariates, presents a key issue in applied statistics \citep[][]{HasTibFri2009}. This issue becomes even more urgent in the context of complex model classes containing various additive predictors. The model-based boosting algorithm thereby automatically leads to sparse solution of the parameter specific structured additive predictors.
Second, fitting regression models via model-based boosting shrinks the effect estimates towards zero. This leads to a decreased variability of predictor effects and typically also to an improvement in the prediction accuracy \citep[][]{MayFenHof2012}. Finally and third, model-based boosting is capable to estimate regression models in high-dimensional settings, for instance, when the number of covariates exceeds the number of observations by far ($p\gg n$), as it frequently occurs in biomedical research \citep[][]{RomEspGot2006, BerPonSpi2015}. It is not easily feasible to apply conventional estimation techniques in these high-dimensional scenarios. As a result, the model-based boosting estimation approach complements existing Bayesian and penalized-likelihood techniques in the context of copula regression models and incorporates valuable modeling features that are of high relevance in applied data analysis.  \\
The paper is structured as follows. We first introduce distributional copula regression models with continuous margins in the section 'Distributional Copula Regression Models'. In the section 'Estimation via Model-Based Boosting' we extend the model-based boosting algorithm to the distributional copula regression class and give insights into the algorithmic structure and hyperparameter tuning. Subsequently, the section 'Simulations' covers extensive simulation studies in order to evaluate the estimation results, the variable selection accuracy and the model building process in low and high dimensional settings and scenarios with dependent and independent responses. In the section 'Analysis of Fetal Ultrasound Data' we apply boosted distributional copula regression to jointly analyze the birth length and weight of fetuses by means of a birth cohort data set. Boosted copula distributional models are of particular interest in this application as the responses are defined on the positive real line and thus require other marginal distributions than the normal distribution. Moreover, the data set consists of 36 covariates, which makes simultaneous variables selection for the different additive predictors of the model parameters an interesting task. Finally, parameter shrinkage induced by boosting in particular suits prediction setups like in delivery management scenarios. We conclude with a thorough discussion and further research ideas concerning boosted copula regression models in the 'Conclusion'.

%% file: sec2.tex
\section{Distributional Copula Regression Models}
\label{sec:BoostRegMods}

In this section, we first review structured additive distributional copula regression models along the lines of \citet{KleKne2016} and introduce specific examples of marginal distributions and copula specifications that are relevant for our simulations and application in the sections 'Simulations' and 'Analysis of Fetal Ultrasound Data'. Subsequently, the concept of structured additive predictors is briefly described. In general, we focus on a bivariate setting, even though the framework of boosting distributional copula regression could be extended to the multivariate case.   

\subsection{The notion of distributional copula regression models}
\label{sec:NotCopMod}
Copulas provide a flexible approach to the construction of multivariate distributions and regression models. As a result of Sklar's theorem, the joint conditional cumulative distribution function (CDF) of the two continuous random response variables $\mY=(Y_1,Y_2)^\top$ given a $p$-dimensional covariate vector $\xvec=(x_1,\ldots,x_p)^\top$ can be expressed as
\begin{equation}
    F_{\mY}(y_{1}, y_{2} \mid \bm{\vartheta}) = C\left(F_{1}\left(y_{1} \mid \bm{\vartheta}^{(1)}\right), F_{2}\left(y_{2} \mid\bm{\vartheta}^{(2)}\right) \mid \bm{\vartheta}^{(c)}\right)
    \label{equ:MultCDF}
\end{equation}
where $\bm{\vartheta} = \left((\bm{\vartheta}^{(1)})^\top, (\bm{\vartheta}^{(2)})^\top, (\bm{\vartheta}^{(c)})^\top\right)^\top\in\dsR^{K_1+K_2+K_c}$ is the vector of model parameters, $F_{1}\left(y_{1} \mid \bm{\vartheta}^{(1)}\right)$ and $F_{2}\left(y_{2} \mid  \bm{\vartheta}^{(2)}\right)$ are the marginal CDFs of the two response variables, respectively, and $C\left(\cdot , \cdot \mid \bm{\vartheta}^{(c)}\right)$ is a uniquely defined copula function \citep{Skl1959, Pat2006}.
The vectors $\bm{\vartheta}^{(\bullet)}$ with $\bullet\in\lbrace 1,2,c\rbrace$ contain the $k=1,\ldots, K_{\bullet}$ parameters $\vartheta_{k}^{(\bullet)}$ of the marginal distributions and the copula function.
In our distributional regression setup all components of $\bm{\vartheta}$ are linked to (potentially different subsets of) the covariate vector $\xvec$ via additive predictors and appropriate link functions (see section 'Structured additive predictors' for details). 
Since $F_{1}\left(y_{1} \mid \bm{\vartheta}^{(1)}\right)$ and $F_{2}\left(y_{2} \mid \bm{\vartheta}^{(2)}\right)$ can be interpreted as uniformly distributed random variables $u_{1}$ and $u_{2}$, respectively, $C\left(\cdot,\cdot \mid \bm{\vartheta}^{(c)}\right)$ is a bivariate distribution function on $[0,1]^2$ that does not depend on the specific marginal CDFs. 
Equation \eqref{equ:MultCDF} also implies the following.  If $C\left(\cdot , \cdot \mid \bm{\vartheta}^{(c)}\right)$ is a conditional copula function and $F_{1}\left(y_{1} \mid \bm{\vartheta}^{(1)}\right)$ and $F_{2}\left(y_{2} \mid \bm{\vartheta}^{(2)}\right)$ are conditional cumulative distributions, then $F_{\mY}$ is a conditional bivariate distribution function with conditional marginal distributions $F_{1}$ and $F_{2}$. Hence, linking together two (potentially different) univariate continuous distributions via a copula function yields a valid bivariate distribution. 
The bivariate conditional probability density function (PDF) is given by 
\begin{equation}
\begin{split}
    f_{\mY}(y_{1}, y_{2} \mid \bm{\vartheta}) = & c\left(F_{1}\left(y_{1} \mid \bm{\vartheta}^{(1)}\right), F_{2}\left(y_{2} \mid  \bm{\vartheta}^{(2)}\right) \bigm\vert \bm{\vartheta}^{(c)}\right) \\
    &\quad f_{1}\left(y_{1} \mid  \bm{\vartheta}^{(1)}\right)  f_{2}\left(y_{2}  \mid \bm{\vartheta}^{(2)}\right), 
    \label{equ:MultPDF}
\end{split}
\end{equation}
where  $f_{1}\left(y_{1} \mid \bm{\vartheta}^{(1)}\right)$ and $f_{2}\left(y_{2} \mid \bm{\vartheta}^{(2)}\right)$ are the marginal PDFs and 
\begin{equation*}
    c\left(F_{1}\left(y_{1} \mid \bm{\vartheta}^{(1)}\right), F_{2}\left(y_{2} \mid \bm{\vartheta}^{(2)}\right) \mid \bm{\vartheta}^{(c)}\right) = \frac{\partial^{2}}{\partial F_{1} \partial F_{2} } F_{1,2}\left(y_{1}, y_{2} \mid \bm{\vartheta}\right)
    \label{equ:copulaPDF}
\end{equation*}
is the copula density. 
For a data set of $n$ bivariate observations $\lbrace (y_{i1},y_{i2})^\top\rbrace_{i=1,\ldots,n}$ with $p$-dimensional covariate vectors $\xvec_1,\ldots,\xvec_n$ and observation-specific distributional parameters $\bm{\vartheta}_{(i)} = \left((\bm{\vartheta}_{(i)}^{(1)})^\top, (\bm{\vartheta}_{(i)}^{(2)})^\top, (\bm{\vartheta}_{(i)}^{(c)})^\top\right)^\top\in\dsR^{K_1+K_2+K_c}$, the density \eqref{equ:MultPDF} induces the joint log-likelihood function
\begin{equation}\label{eq:loglik}
\begin{split}
    l(\varthetavec_{(1:n)}) \equiv & \sum_{i = 1}^{n} \log\left(c\left(F_{1}\left(y_{1i}  \mid \bm{\vartheta}_{(i)}^{(1)}\right), F_{2}\left(y_{2i} \mid \bm{\vartheta}_{(i)}^{(2)}\right) \mid \bm{\vartheta}_{(i)}^{(c)}\right)\right) + \\
    &\quad \sum_{i = 1}^{n} \sum_{d\in\lbrace 1,2\rbrace} \log\left(f_{d}\left(y_{di} \mid \bm{\vartheta}_{(i)}^{(d)}\right)\right),
\end{split}
\end{equation}
where $\varthetavec_{(1:n)}=\left(\varthetavec_{(1)}^\top,\ldots,\varthetavec_{(n)}^\top\right)^\top\in\dsR^{n\times(K_1+K_2+K_c)}$. We introduce this notation to differentiate the $K_{\bullet}$-dimensional vectors $\bm{\vartheta}_{(i)}^{(\bullet)}$ (all distributional parameters of one of the margins or the copula parameter for one specific observation $i$ in the data) from the $n$-dimensional vectors $\varthetavec_k^{(\bullet)}=(\vartheta_{k,(1))}^{(\bullet)},\ldots,\vartheta_{k,(n))}^{(\bullet)})^\top$ (one specific distributional parameter of the joint distribution for all $n$ observations).

\subsection{Examples of marginal distributions}

\label{sec:MargDist}
Next, we present two important marginal distributions, namely the Log-Normal and the Log-Logistic distributions, which are considered in section 'Simulations' and section 'Analysis of Fetal Ultrasound Data'. Both distributions have two distributional parameters, which we denote as $\mu$ and $\sigma$.

\subsubsection{Log-Normal}
\label{sec:LogNormDist}
A continuous random variable $Y$ is log-normally distributed with location  $\mu\in\dsR$ and scale $\sigma\in\dsR^+$ if $X=\log(Y)$ has a Normal distribution with mean $\mu$ and standard deviation $ \sigma$. 
Its PDF and the CDF are given by
\begin{equation*}\begin{aligned}
        f(y \mid \mu, \sigma) &= \frac{1}{\sqrt{2 \pi \sigma^{2}}} \frac{1}{y} \exp\left( - \frac{(\log(y) - \mu)^{2}}{2\sigma^{2}}\right)\\
        \text{and}\\
          F(y \mid \mu, \sigma) &= \Phi \left(\frac{\log(y) - \mu}{\sigma}\right).
    \label{equ:PDFLOGNO}
\end{aligned}\end{equation*}

\subsubsection{Log-Logistic}
\label{sec:LogLogDist}
The Log-Logistic distribution is a special case of the four-parametric Generalized Beta Prime distribution where the first two shape parameters are set equal to $1$. For a  Log-Logistic distribution with scale parameter $\mu\in\dsR^+$ and shape parameter $\sigma\in\dsR^+$ the PDF and the CDF can be expressed as 
\begin{equation*}\begin{aligned}
        f(y \mid \mu, \sigma) &= \frac{\sigma \left(\frac{y}{\mu} \right)^{\sigma}}{y \left(1+ \left(\frac{y}{\mu} \right)^{\sigma} \right)^{2}} \\
        \text{and} \\
        F(y \mid \mu, \sigma) &= \frac{1}{1 + \left( \frac{y}{\mu} \right)^{-\sigma}}.
    \label{equ:PDFLOGLOG}
\end{aligned}\end{equation*}
In the section 'Analysis of Fetal Ultrasound Data', we consider the mean of the Log-Logistic distribution, which is defined as 
\begin{equation}\label{eq:meanloglog}
        E(Y) = \frac{\frac{\mu \pi}{\sigma}}{\sin\left(\frac{\pi}{\sigma}\right)}.
\end{equation}

\subsection{Dependence structure}
\label{sec:DepStruc}
\begin{figure}[ht]
\includegraphics[width=1\textwidth]{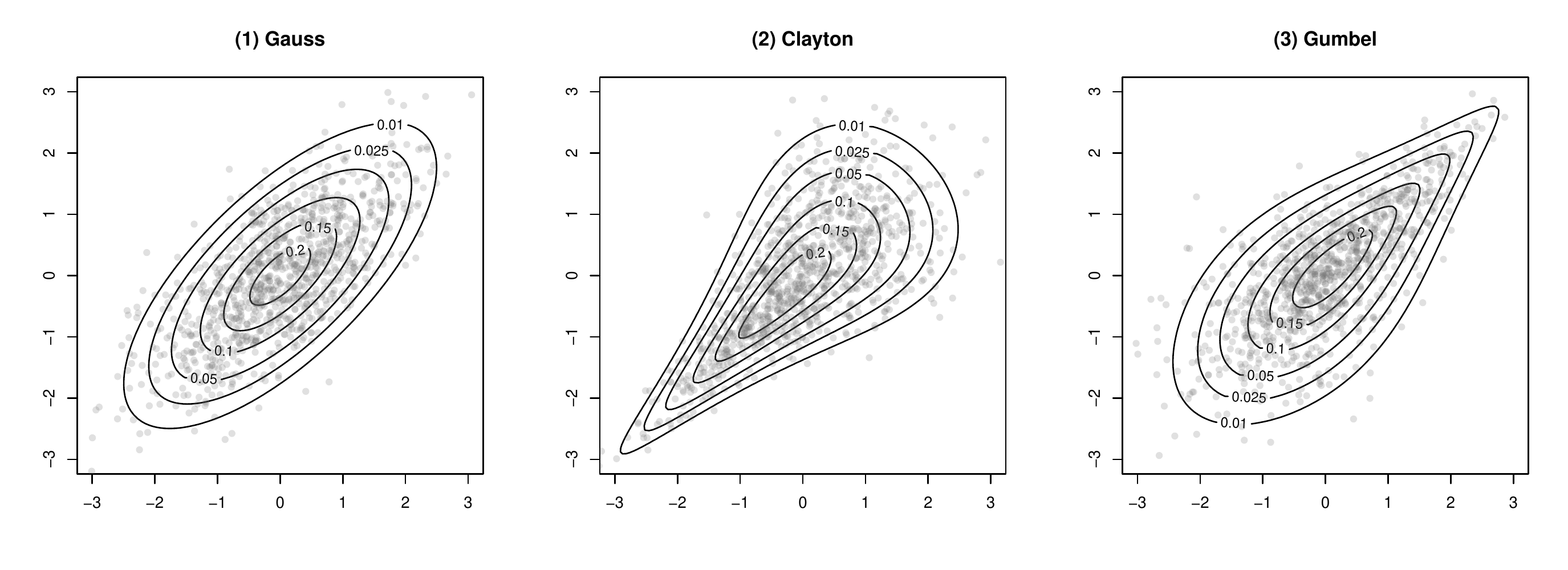} 
\caption{Contour lines of densities and 1000 simulated observations for the Gaussian, the Clayton and the Gumbel copula with standard normal marginal distributions. The association parameters are set to 0.71, 2, 2, respectively, which induce a medium positive correlation. The Gaussian copula has no tail dependence. The Clayton and the Gumbel copula exhibit a strong lower and upper tail dependence, respectively. The distributions and samples are generated via the \texttt{VineCopula} package \citep{NagSchSto2021}.}
\label{fig:ContCopPlots}
\end{figure}

Archimedean and elliptical copulas are popular copula function choices in practice \citep[][]{McNFreEmb2005} and allow for different dependence scenarios as presented in Figure \ref{fig:ContCopPlots}.
To summarized the associations between the component's of a multivariate response beyond the commonly employed linear Pearson's correlation coefficient, important dependence measures in a copula context are given by the Kendall's tau rank correlation $\tau^\kappa$ and the upper and the lower tail dependence, which are defined as $\lambda_{u} = \lim_{q \to 1^{-}} P(y_{2} > F_{2}^{-1}(q) \mid y_{1} > F_{1}^{-1}(q))$ and $\lambda_{l} = \lim_{q \to 0^{+}} P(y_{2} \leq F_{2}^{-1}(q) \mid y_{1} \leq F_{1}^{-1}(q))$, respectively \citep[][]{McNFreEmb2005}. The most important example of elliptical copulas is the Gaussian copula \citep{Song2000}, which has no tail dependence, i.e.~$y_{1}$ and $y_{2}$ are independent in the limit and $\lambda_u=\lambda_l=0$. In contrast, Archimedean copulas do allow for lower or upper tail dependence, depending on the specific choice. The Clayton and Gumbel copulas are prominent members of this family, the former allowing for lower tail dependence and the latter for upper tail dependence.  
Their counter-clockwise rotated versions enable the modeling of negative dependence \citep[][]{BreSch2013}. Note that all introduced copula functions contain only one parameter of dependence, which we denote as $\vartheta^{(c)}$. Yet, the framework of boosting distributional copula regression can also be extended to copulas with more parameters. 
An overview of the introduced copulas and the respective formulas for Kendall's tau and the lower and upper tail dependence are given in Table \ref{tab:TabCop}. More details on theoretical properties \citep{Joe1997,Nel2006,Joe2014} and conditional copula models \citep{Pat2002, VatCha2015} are discussed in the corresponding literature. 
\begin{table}[h]
\caption{\label{tab:TabCop} Details on the  Gaussian, Clayton and Gumbel copulas. The functions $\Phi(\cdot)$ and $\Phi_{2}\left(\cdot, \cdot; \theta \right)$ denote the CDFs of the standard univariate normal distribution and bivariate normal distribution with standard normally distributed margins and linear correlation parameter $\rho\equiv\vartheta^{(c)}$, respectively. For the Gumbel copula we introduced $v_1=(-\log(u_{1}))^{\vartheta^{(c)}}$ and $v_2=(-\log(u_{2}))^{\vartheta^{(c)}}$.
}
\small
\begin{center}
\begingroup
\setlength{\tabcolsep}{6pt} 
\renewcommand{\arraystretch}{1.5}
\begin{tabular}{ l l l l }
\hline
 Copula & $C\left(u_{1},  u_{2}; \vartheta^{(c)}\right)$ & Tail dependence & $\tau^\kappa$\\
 \hline
Gaussian & $\Phi_{2}\left(\Phi^{-1}(u_{1}), \Phi^{-1}(u_{2}); \vartheta^{(c)}\right)$ & $\lambda_u=\lambda_l=0$ &  $\frac{2}{\pi} \text{arcsin}(\vartheta^{(c)})$\\  
Clayton & $\left(u_{1}^{-1} + u_{2}^{-1} - 1\right)^{-1/\vartheta^{(c)}}$ & $\lambda_u=0,\;\lambda_{l} = 2^{\tfrac{-1}{\vartheta^{(c)}}}$  & $\frac{\vartheta^{(c)}}{(\vartheta^{(c)} + 2)}$\\
Gumbel &
$\begin{aligned}[t]
   \exp\left\lbrack -\left( v_1 
   + v_2 \right)^{1/\vartheta^{(c)}}\right\rbrack
\end{aligned}$ 
& $\lambda_{u} = 2 - 2^{\tfrac{1}{\vartheta^{(c)}}},\;\lambda_u=0$ & $1 -\frac{1}{\vartheta^{(c)}}$\\
 \hline
\end{tabular}
\endgroup
\end{center}
\end{table}

\subsection{Structured additive predictors}
\label{sec:AddPred}
In distributional copula regression, each distributional parameter $\vartheta_{k}^{{(\bullet)}}$ can be associated with a structured additive predictor $\eta_{k}^{{(\bullet)}}$ through appropriate monotonic response functions $h_{k}^{(\bullet)}(\cdot)$ to ensure the restrictions on the respective parameter spaces. This is similar to GAMs, where link functions are used to model the conditional expectation of a distribution from the exponential family, which leads to $\vartheta_{k}^{(\bullet)} = h_{k}^{(\bullet)}\left(\eta_{k}^{{(\bullet)}}\right) \quad\mathrm{and}\quad \eta_{k}^{{(\bullet)}} = \left(h_{k}^{(\bullet)}\right)^{-1} \left(\vartheta_{k}^{(\bullet)}\right), $ where the inverse $\left(h_{k}^{(\bullet)}\right)^{-1}$ is the corresponding link function. 
For instance, the location parameter $\mu$ of the Log-Normal distribution can be directly set equal to its additive predictor $\eta_{\mu}$, i.e.~$h_{\mu}$ is the identity function. For the scale parameter $\sigma$ of the Log-Normal (corresponding to the standard deviation) and both parameters of the Log-Logistic distribution we use the $\log(\cdot)$ link function to guarantee the restriction on the positive real line. The link functions of the Gaussian, the Clayton and the Gumbel copula dependence parameters are given by $\text{tanh}^{-1}(\eta^{(c)})$ with $\vartheta^{(c)} \in (-1,1)$, $\log(\eta^{(c)})$ with $\vartheta^{(c)} \in (0,\infty)$ and  $\log(\eta^{(c)} - 1)$ with $\vartheta^{(c)} \in (1,\infty)$, respectively. 
Now, corresponding to the GAMLSS class each structured additive predictor is modelled through a sum of an overall intercept term $\beta_{0, k}^{(\bullet)}$ plus $J_{k}^{(\bullet)}$ generic functions $f_{j,k}^{(\bullet)}(\xvec_{j})$ expressed as
\begin{equation*}
    \eta_k^{(\bullet)} = \beta_{0,k}^{(\bullet)} + \sum_{j = 1}^{J_k^{(\bullet)}} f_{j,k}^{(\bullet)}(\xvec_{j}),
    \label{equ:GenAddPred}
\end{equation*}
where each parameter specific predictor is associated to an individual subset of covariates, i.e.~$\xvec_{1}, \dots, \xvec_{J_{k}^{(\bullet)}}$.
The intercept is the overall level of the predictor when all function evaluations are zero and the effects of the covariates on the parameter models are determined by the functions $f_{j,k}^{(\bullet)}(\xvec_{j})$.
For instance, in the application in the section 'Analysis of Fetal Ultrasound Data' we fit models that incorporate linear and non-linear effects. 
Dropping the parameter superscripts $(\bullet)$ and parameter index $k$ a single generic linear effect is given by a function of the form $f_{j}^{linear}(x_{j}) = \beta_{j} x_{j}$. 
Non-linear effects of univariate covariates $x_j$ can be modeled via P-splines with $L_{j}$ B-Spline basis functions $B_{j, l_{j}}(x_{j})$ resulting in the generic smooth function $f_{j}^{smooth}(x_{j}) = \sum_{l_{j} = 1}^{L_{j}} \beta_{j, l_{j}} B_{j,l_{j}}(x_{j})$.
A second order difference penalty on the coefficients is introduced to control for the smoothness of the non-linear effect \citep[][]{EilMar1996}. 
Many types of smoothing functions, such as smoothing or regression splines, exist \citep{Woo2017}. 
Moreover, further covariate effect types like individual or group specific random effects or spatial effects can be incorporated in the structured additive predictors \citep{FahKneLan2013}.
In summary, each model parameter specific structured additive predictor is related to an individual set of covariates, which are also based on different functional forms.

%% file: sec3.tex
\section{Estimation via Model-Based Boosting}\label{sec:boosting}

Our proposed method includes the estimation of distributional copula regression models via model-based boosting. This section provides an overview on the algorithm and covers questions related to hyperparameter tuning and model building. 

\subsection{Algorithm for boosting distributional copula  models}
Parameter estimation of the distributional copula regression model of a bivariate continuous random variable $\mY$  can be expressed in terms of the optimization problem 
\begin{equation*}
    \hat{\bm{\eta}} = \argmin_{\bm{\eta}} \left\{E_{\bm{Y}} \left[\omega\left(\bm{Y}; \eta_{1}^{(1)}, \dots, \eta_{K_{1}}^{(1)}, \eta_{1}^{(2)}, \dots, \eta_{K_{2}}^{(2)}, \eta_{1}^{(c)}, \dots, \eta_{K_{c}}^{(c)} \right) \right]\right\},
\end{equation*}
where $\bm{\eta} = \left(\eta_{1}^{(1)}, \dots, \eta_{K_{1}}^{(1)}, \eta_{1}^{(2)}, \dots, \eta_{K_{2}}^{(2)}, \eta_{1}^{(c)}, \dots, \eta_{K_{c}}^{(c)}\right)^\top \in\dsR^{K_1+K_2+K_c}$ is the vector of additive predictor functions of the model parameters, the respective estimates of the additive predictor functions are given by $\bm{\hat{\eta}} = \left(\hat{\eta}_{1}^{(1)}, \dots, \hat{\eta}_{K_{1}}^{(1)}, \hat{\eta}_{1}^{(2)}, \dots, \hat{\eta}_{K_{2}}^{(2)},
\hat{\eta}_{1}^{(c)}, \dots, \hat{\eta}_{K_{c}}^{(c)}\right)^\top \in\dsR^{K_1+K_2+K_c}$ and $\omega(\cdot)$ denotes the loss function. 
Considering a data sample with observations $i = 1,\dots,n$, we go over to minimize the empirical risk 
\begin{equation*}
    \frac{1}{n} \sum_{i = 1}^{n} \omega\left(\bm{y}_{i}; \eta_{(i),1}^{(1)}, \dots, \eta_{(i),K_{1}}^{(1)}, \eta_{(i),1}^{(2)}, \dots, \eta_{(i),K_{2}}^{(2)},
    \eta_{(i),1}^{(c)}, \dots, \eta_{(i),K_{c}}^{(c)}\right)
    \label{equ:EmpRisk}
\end{equation*}
over $\bm{\eta}_{(i)} = \left(\eta_{(i),1}^{(1)}, \dots, \eta_{(i),K_{1}}^{(1)}, \eta_{i1}^{(2)}, \dots, \eta_{(i),K_{2}}^{(2)}, \eta_{i1}^{(c)}, \dots, \eta_{(i),K_{c}}^{(c)}\right) \in\dsR^{K_1+K_2+K_c}$ instead. The loss function $\omega(\cdot)$ measures the discrepancy between the observed responses and the estimated additive predictors. In a regression context, a common choice for $\omega$ is the negative log-likelihood of the (bivariate) response distribution, which is provided in Equation \eqref{eq:loglik} for distributional copula regression models.\\ 
The fundamental idea of model-based gradient boosting is to sequentially minimize the empirical risk by a stepwise descent of the loss function's gradient in function space. In more detail, the algorithm starts with an initial specification of the additive predictors and a set of pre-defined regression-like base-learners, e.g.~simple linear models or regression splines with low degrees of freedom, for every model parameter. Note that the type of base-learner determines the type of effect the covariates have on the respective model parameter. In every iteration $m$ the parameter-specific sets of base-learners are fitted one-by-one to the current respective negative partial gradient vectors (derivatives of the loss with respect to the different model parameter predictors $\eta_{k}^{(\bullet)}$). Subsequently, the best base-learner in terms of a residual sum of squares criterion is selected for each parameter $\vartheta^{(\bullet)}_{k}$. Two algorithmic versions for boosting distributional regression models exist, namely the \textit{cyclic} \citep{MayFenHof2012} and the \textit{noncyclic} \citep{ThoMayBis2018} algorithm. The \textit{cyclical} version updates each additive predictor function with the respective best fitting base-learner in each iteration successively, using the other estimates as offset values. The \textit{noncyclic} version incorporates an additional selection step of the best-fitting model parameter. More specifically, in each iteration only the parameter model that yields the highest overall loss reduction is updated with its parameter specific best base-learner. A strictly additive aggregation of the best fitting base learners over the course of the boosting procedure accounts for the additive structure of the resulting predictor functions (see section 'Structured additive predictors') in both algorithmic versions. For reasons of faster tuning properties, we focus on the \textit{noncyclic} algorithm in the following and summarize the main algorithm in a generic way in Algorithm \ref{alg:the_alg}.
\begin{algorithm}[htbp]
\small
  \caption{Component-wise gradient boosting algorithm for copula regression} 
  \vspace{4mm}
  \textbf{Initialization} 
  \vspace{2mm}
  \\
  For each $\vartheta^{(\bullet)}_{k} \in \bm{\vartheta}$:
  \begin{enumerate}
      \item[1.] Initialize $\hat{\eta}^{(\bullet)}_{k}$ with an offset value $\hat{\eta}^{(\bullet)}_{k, [0]}$, i.e.~with zero
      \item[2.] Define $\left(b^{(\bullet)}_{k,1}(x_{1}), \dots, b^{(\bullet)}_{k,p}(x_{p})\right)$, i.e.~a model parameter specific set of base learners where each covariate is associated with one base learner $b^{(\bullet)}_{k,j}(x_{j})$ with $j = 1,\dots,p$
  \end{enumerate}
  \vspace{2mm}
  \textbf{Boosting}
  \vspace{1mm}
  \\
  For $m = 1$ to $m_{\mbox{\scriptsize{stop}}}$:
  \vspace{1mm}
    \begin{enumerate}
      \item[3.] For each $\vartheta^{(\bullet)}_{k} \in \bm{\vartheta}$:
      \begin{enumerate}
        \item[(a)] Compute the current parameter-specific negative gradient vector $-\bm{g}^{(\bullet)}_{k, [m]}$ as 
        $$
        -\bm{g}^{(\bullet)}_{k, [m]}= - \left(g_{k, [m]}^{(\bullet)}(\bm{x}_{i})\right)_{i =1,\dots,n} = - \left( \left. \frac{\partial \omega\left(\bm{y}_{i}; {\hat{\etavec}_{(i)}}\right)}{\partial \eta^{(\bullet)}_{k}} \right\vert_{\bm{\hat{\eta}}_{(i)} = \bm{\hat{\eta}}_{[m-1]}(\bm{x}_{i})}\right)_{i=1,\dots,n}
        $$
        \item[(b)] Fit $-\bm{g^{(\bullet)}}_{k, [m]}$ to each parameter-specific base learner $b^{(\bullet)}_{k,j}(x_{j})$. 
        \item[(c)] Select the best fitting base learner $\hat{b}^{(\bullet)}_{k, j^{*}}$ by the residual sums of squares criterion
        $$
        j^{*} = \argmin_{j \in 1, \dots, p} \sum_{i = 1}^{n} \left( - g_{k, [m]}^{(\bullet)}(\bm{x}_{i}) - \hat{b}^{(\bullet)}_{k,j}(x_{ij})\right)^{2}
        $$
        \item[(d)] Compute the potential improvement of an update of the parameter-specific additive predictor with $\hat{b}^{(\bullet)}_{k, j^{*}}$ in terms of loss reduction
        $$
        \Delta \omega_{\vartheta^{(\bullet)}_{k}} = \sum_{i = 1}^{n} \omega \left( \bm{y}_{i}; \hat{\eta}^{(\bullet)}_{k, [m-1]}(\bm{x}_{i}) + s \cdot \hat{b}^{(\bullet)}_{k, j^{*}}(x_{ij})\right)
        $$
      \end{enumerate}
      \item[4.] Update only the parameter model, which yields the highest loss reduction, i.e. 
      $\vartheta^{(\bullet)*}_{k} = \argmin_{\vartheta^{(\bullet)}_{k} \in \bm{\vartheta}} (\Delta \omega_{\vartheta^{(\bullet)}_{k}})$, with the respective best fitting base learner
      $$
      \hat{\eta}_{k, [m]}^{(\bullet)*}(\bm{x}_{i}) = \hat{\eta}_{k, [m-1]}^{(\bullet)*}(\bm{x}_{i}) + s \cdot \hat{b}^{(\bullet)}_{k, j^{*}}(x_{j})
      $$
      \item[5.] For the other parameter models, i.e. $\vartheta^{(\bullet)}_{k} \neq \vartheta^{(\bullet)*}_{k}$, set
      $$
      \hat{\eta}^{(\bullet)}_{k, [m]}(\bm{x}_{i}) = \hat{\eta}^{(\bullet)}_{k, [m-1]}(\bm{x}_{i})
      $$
      
  \end{enumerate}
\label{alg:the_alg}  
\end{algorithm}
\subsection{Choice of hyperparameters}
Two hyperparameters of the boosting algorithm are crucial for the parameter estimation and the complexity of the additive predictors, namely the step length $s$ and the stopping iteration $m_{\mbox{\scriptsize{stop}}}$. The step length is involved in the additive updates, adding only a small proportion of the base-learner fit to the models. This guarantees the stability of the boosting algorithm and shrinks the parameter estimates to zero in combination with an appropriate stopping criterion. The stopping iteration $m_{\mbox{\scriptsize{stop}}}$ defines the iteration after which further updates are not longer necessary. Early stopping, i.e. when the algorithm is stopped before convergence, automatically controls the model complexity of the parameter specific additive predictors. Particularly, component-wise gradient boosting constitutes an intrinsic variable selection mechanism by assigning exactly one base learner to each covariable $x_{1}, \dots, x_{p}$ for each model parameter \citep[][]{BuhYu2003, MayFenHof2012}. Consequently, in each iteration only the most informative variable is selected. Stopping the algorithm early finally leads to the inclusion of the most-informative variables in the additive predictors, while the less informative covariates are dismissed. \\
Tuning is the task of finding appropriate values for the hyperparameters. In the model-based boosting framework, the optimal $m_{\mbox{\scriptsize{stop}}}$ is usually determined via cross validation \citep[][]{HotLeiZei2005, BuhHot2007, MayFenHof2012}. More precisely, the original data is split into a training and test fold several times. While a boosting model is fitted to the training sets, the test folds are used to assess the performance of the model by means of the predictive risk. Eventually, the optimal $m_{\mbox{\scriptsize{stop}}}$ is assigned to the iteration of the smallest predictive risk averaged over the  different test folds. The step length $s$ might be set to a fix small value (much less than 1) in advance \citep[][]{MayFenHof2012}. The simulation studies from section 'Simulations' suggest that a step length of $0.01$ is suitable for one parametric conditional copula models with two parametric marginal distributions. 

\subsection{Model building} \label{sec:ModBuild}
Copula regression offers an extremely flexible framework to jointly model response variables with potentially different marginal distribution conditional on covariate information. In practice, the researcher is confronted with the choice of the marginal distributions and the copula function. In a penalized likelihood-based context, the Akaike information criterion (AIC) and/or Bayesian information criterion (BIC) in combination with normalized quantile residuals is proposed for those model building decisions \citep{MarRad2017}. The boosting algorithm yields regularized fits, which makes the evaluation of residuals difficult as \citet{HofMay2016} describe. As a consequence, for boosted univariate regression models the authors suggest to consider the predictive risk, i.e. the risk on a new data set, to make decisions on the appropriate distribution function. We follow their approach for the decision concerning the marginal distributions and expand the concept of the predictive risk also to the choice of the appropriate copula function.

%% file: sec4.tex
\section{Simulations}
\label{sec:sim}

Estimating regression models via component-wise gradient boosting yields some interesting properties like parameter shrinkage, an intrinsic variables selection mechanism and the capability to fit models in high-dimensional data settings. The following simulation study investigates these properties in the context of copula regression with continuous margins. More specifically, we are interested in answering four questions, which are crucial for the application of boosted copula regression in real data situations:
\begin{enumerate}
    \item[(a)] Does the boosting algorithm estimate the correct (but shrunk) effects of the informative covariates on the different parameter models?
    \item[(b)] Is the boosting algorithm able to separate the informative from the non-informative covariates? 
    \item[(c)] In case of independence between the response variables, does the boosting algorithm dismiss the covariates in the dependence parameter model?
    \item[(d)] Is the correct copula model selected by the predictive risk in the presence of misspecified models? Hence, is the predictive risk a valid tool to choose an appropriate copula function?
\end{enumerate}
We do not compare the results of boosting to competitors like the penalized-likelihood based copula regression method implemented in the \texttt{GJRM} package \citep{MarRad2017} as our estimates are naturally shrunk and for \texttt{GJRM} it is not feasible to fit models in high-dimensional settings. Yet, for a low-dimensional example with only linear covariate effects we confirm that our algorithm converges to the estimates of the \texttt{GJRM} implementation for all three copulas. The respective results and plots are provided in the Appendix (A.1).

\subsection{Simulation design}

We carry out simulations that cover a low ($p = 20$) and a high-dimensional ($p = 1000$) data setup for the Gaussian, the Clayton and the Gumbel copula in combination with the Log-Normal and the Log-Logistic marginal distributions. 
The number of observations is set to $n = 1000$ and each simulation scenario is repeated $100$ times. The covariates are sampled independently from a uniform distribution on $(-1, 1)$. Each model parameter is associated with a subset of informative covariates in a linear or non-linear fashion. More specifically, the parameter models are given by
\begin{equation*}\begin{aligned}
    h_{\mu_{1}}(\mu_{1}) &= -0.75  x_{1} + 0.5  \cos(\pi  x_{3}) \\
    h_{\mu_{2}}(\mu_{2}) &= 0.5 - 0.7  x_{1}  - 0.02\exp(2(x_{2} + 1)) \\
    h_{\sigma_{1}}(\sigma_{1}) &= -0.7 + 0.5  \sin(\pi  x_{3}) \\
    h_{\sigma_{2}}(\sigma_{2})& = 2 + 0.5 x_{2} \\
    h_{\vartheta^{(c)}}(\vartheta^{(c)}) &= -0.8 + 1.5  \log(4.5 - 1.7 \sin(\pi  x_{4})).
\end{aligned}\end{equation*}
Note that in our specification $x_{1}, \dots, x_{4}$ are informative covariates, while $x_{5}, \dots, x_{p}$ have no true effect on the parameter models. All simulations are carried out in the programming language  \texttt{R} \citep{RCorTea2020}. The artificial data sets for the three copulas in combination with the marginal distributions are generated by means of the R packages \texttt{copula} \citep{Yan2007} and \texttt{gamlss} \citep{RigSta2005}.\\
We choose cubic P-splines with 20 equidistant knots, a second-order difference penalty and 4 degrees of freedom as base learners. Following the component-wise boosting concept introduced in the section 'Estimation via Model-Based Boosting' one P-spline base learner is defined for each covariate $p$ and every distribution parameter model. Therefore, in every iteration $m$ the boosting algorithm selects from 100 and 5000 different base learners in the low- and the high-dimensional setting, respectively. In order to determine the optimal $m_{\mbox{\scriptsize{stop}}}$, we evaluate the empirical risk of an additional independent and identically distributed data set with 1500 observations that follows the same distribution as the original data \citep[][]{MayFenHof2012}. 
The copula models are estimated via the noncyclic model-based boosting implementation of the R package \texttt{gamboostLSS} \citep{ThoMayBis2018}.

\subsection{Results for the low-dimensional setting}

\subsubsection*{Parameter estimation}

\begin{figure}[htbp]
\includegraphics[width=0.95\textwidth]{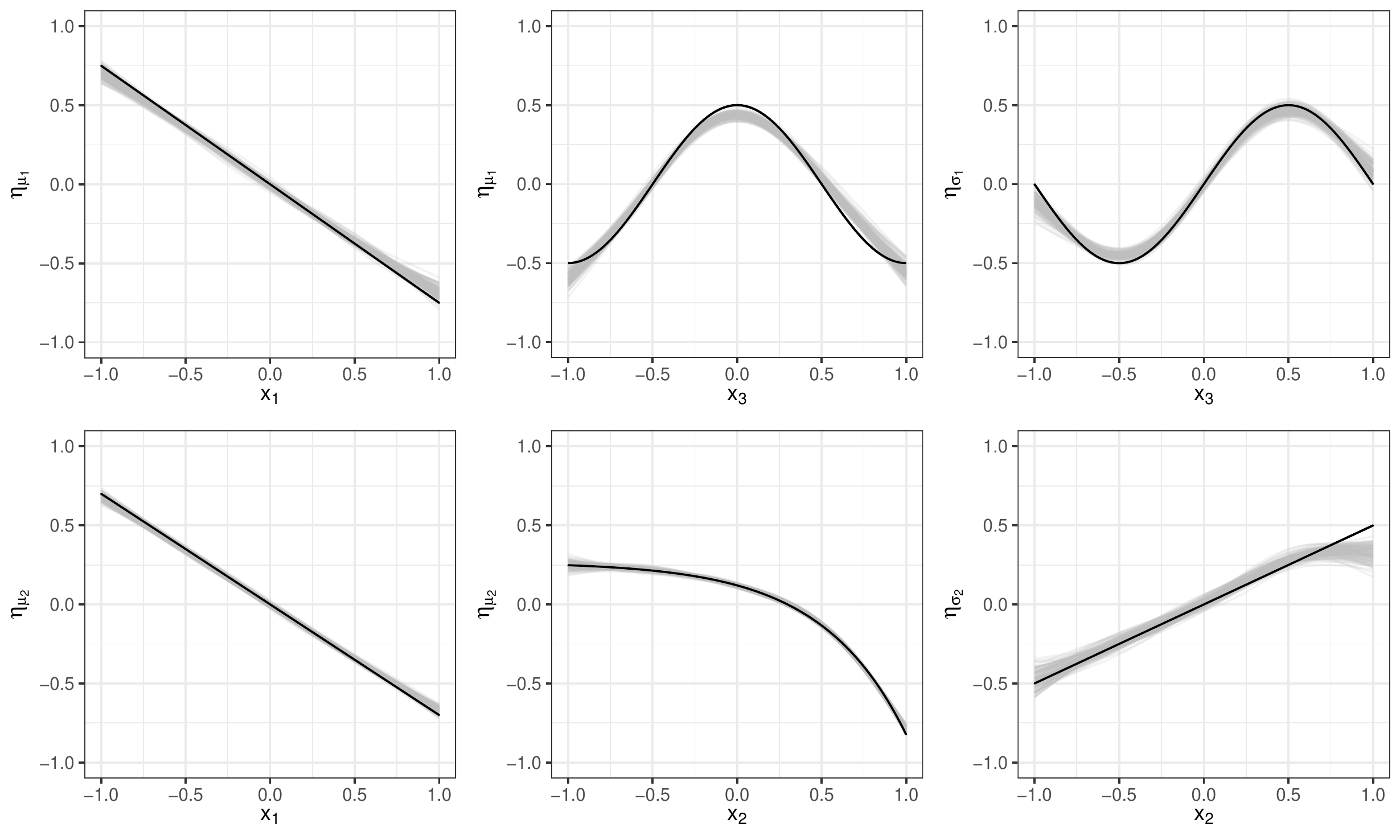} 
\caption{Effect estimates of the marginal models for the Gaussian copula in the low-dimensional setting. The first and the second row of plots show the effect estimates for informative covariates on the parameters of the first and the second marginal distribution (i.e.~Log-Normal and Log-Logistic distribution), respectively. In each plot the gray lines represent the estimates of the 100 runs while the black line displays the original effect.}
\label{fig:LowDimMargEst}
\end{figure}

\begin{figure}[htbp]
\includegraphics[width=0.95\textwidth]{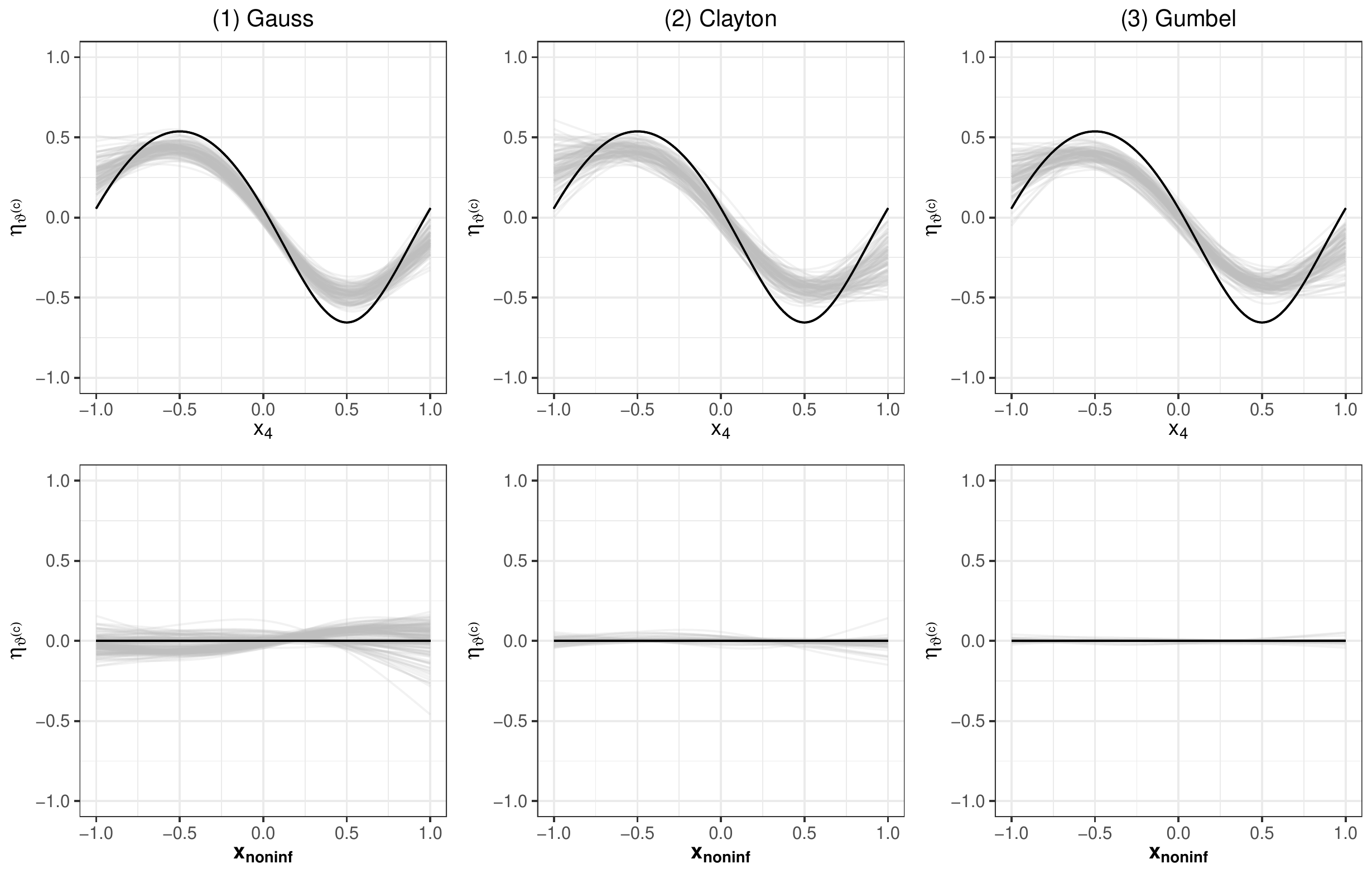} 
\caption{Dependence parameter fits for the Gaussian, Clayton and Gumbel copula in the low-dimensional setting. The first and second row of plots shows the fits of the informative and the non-informative covariate(s) on the copula parameter, respectively. In each plot the gray lines represent the estimates of the 100 runs while the black line displays the original effect.}
\label{fig:LowDimCopEst}
\end{figure}

Figure \ref{fig:LowDimMargEst} shows the estimated effects of the informative covariates on the marginal distribution parameters for the Gaussian copula fit.
The effects on the marginal distribution parameters for the Clayton and the Gumbel copula fit essentially yield the same results and are given in the Appendix (A.2.1). The boosting estimates reflect the true structure of the  informative covariates on each parameter of the marginal models. Note that the estimates are slightly shrunk towards zero, though, which is a result of the early stopping of the algorithm. The plots of the estimates of the non-informative covariates, if they were selected, can be found in the Appendix (A.2.2). As expected, the estimates of the non-informative covariates generally gather around zero.
The impact of the covariates on the copula dependence parameters of the three copulas are presented in Figure \ref{fig:LowDimCopEst}. For all three copulas the true structure of the informative covariate effects is estimated correctly. However, in comparison to the marginal fits the estimates are substantially more shrunk and have a higher variance. For the Gaussian copula the estimates of the non-informative covariates on the dependence parameter show a slight deviation from zero. This is not the case for the Clayton and Gumbel fits.

\subsubsection*{Variable selection performance}
In each simulation run the informative variables were correctly selected for every model parameter, resulting in correct selection rates of 100 \%. Table \ref{tab:LowDimSelAcc} provides the average selection rates of the non-informative covariates for each model parameter and the three copula functions. In general, the $\mu$ models tend to select many non-informative variables. More precisely, with average selection rates of 96.94 \% and 97.28 \% almost all non-informative variables are selected for the $\mu_{2}$ models of the Gaussian and the Clayton copula, respectively. Considering the Gumbel copula, the $\mu_{1}$ model shows with 83.44 \% a high average selection rate of the non-informative covariates. The selection mechanism in the copula dependence parameters performs reasonably well with average selection frequencies of 7.63 \%, 2.11 \% and 1.16 \% for the Gaussian, Clayton and Gumbel copula, respectively. Note that the dependence parameter model of the Gaussian copula includes the most non-informative variables. This matches our previous observation of a stronger impact of non-informative covariates on the dependence parameter in the Gaussian case from Figure \ref{fig:LowDimCopEst}.
\begin{table}[htbp]
\caption{Selection rates of the non-informative covariates averaged over the 100 simulation runs for the different model parameters and copula functions in the low-dimensional setting.}
\centering
\begin{tabular}{ |l|l|l|l| }
\hline
Model parameter & Gauss &  Clayton  & Gumbel\\
\hline
$\mu_{1}$ & 32.06\% & 58.39\% & 83.44\% \\
\hline
$\mu_{2}$ & 96.94\% & 97.28\% & 61.33\% \\
\hline
$\sigma_{1}$ & 19.00\% & 33.58\% & 46.42\% \\
\hline
$\sigma_{2}$ & 17.84\% & 21.31\% & 31.26\% \\
\hline
$\vartheta^{(c)}$ & 07.63\% & 02.11\% & 01.16\% \\
\hline
\end{tabular}

\label{tab:LowDimSelAcc}
\end{table}

\subsubsection*{Estimation with independent responses}
We also fit the three copula models to data that follows the same marginal distributions but has independent responses. 
In this scenario the covariates are not informative for the copula dependence parameters $\vartheta^{(c)}$ and we expect that their structured additive predictor models do not integrate any covariates over the course of the fitting procedure. \\ 
For the low-dimensional setting the Clayton and the Gumbel copula do not select any covariates in their copula parameter models. In case of the Gaussian copula 1.83 variables are integrated on average.

\subsubsection*{Predictive risk}
Finally, the performance of the predictive risk as a tool to select the correct copula function is evaluated. 
\begin{figure}[htbp]
\includegraphics[width=0.95\textwidth]{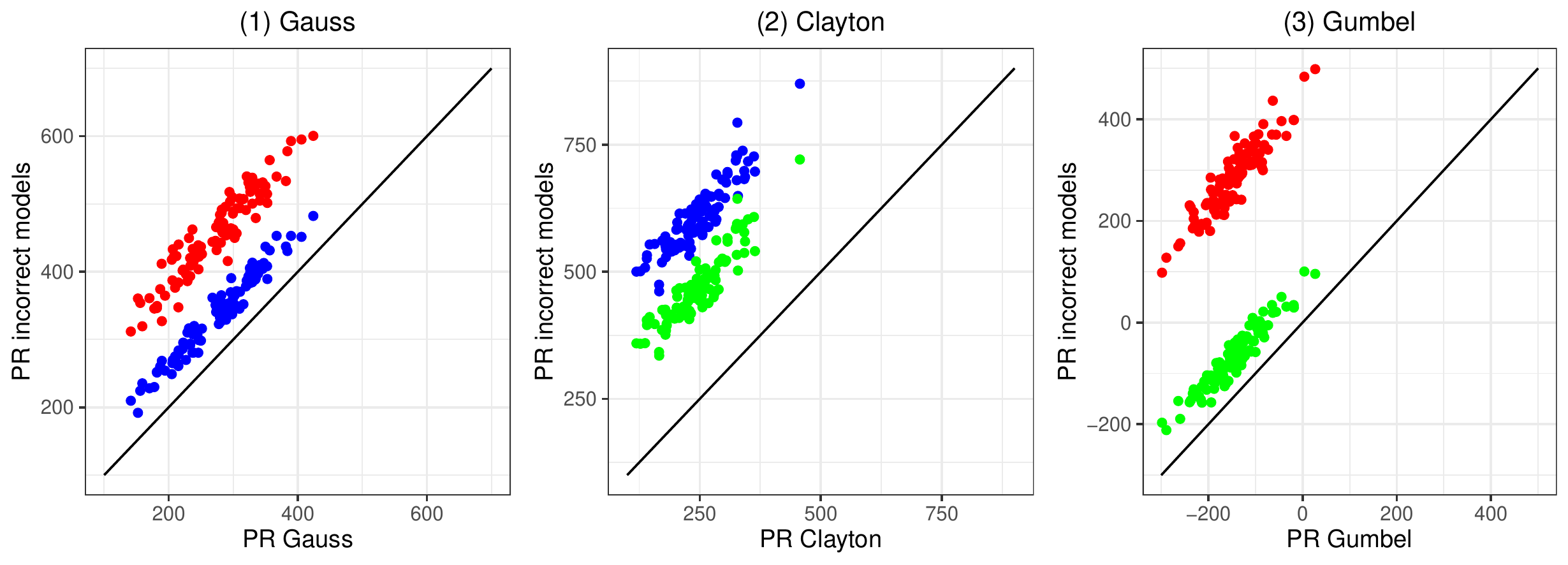} 
\caption{Predictive risk (PR) values of the correct model (x-axis) versus the incorrect models (y-axis) in the low-dimensional data case. For (a), (b) and (c) the Gaussian, the Clayton and the Gumbel copula is the correct model, respectively. The red points mark the comparison to the incorrect Clayton copula, the blue points to the incorrect Gumbel copula and the green points to the incorrect Gaussian copula.}
\label{fig:LowDimPR}
\end{figure}
Figure \ref{fig:LowDimPR} displays the values of the predictive risk for the correct copula specification on the x-axis versus the values of the predictive risk of the misspecified models on the y-axis for each copula function. All points for all three correct copulas are located above the diagonal line, indicating that the true model is always selected. Note that for a correct selection the predictive risk of the true model needs to be smaller than the predictive risk value of the misspecified models. The predictive risk values of the Clayton and the Gumbel copula yield the greatest differences, which can be explained by their different tail dependencies. Moreover, the Gaussian and the Gumbel copula in general show the smallest differences in terms of the predictive risk.

\subsection{Results for the high-dimensional setting}

\subsubsection*{Parameter estimation}

\begin{figure}[htbp]
\includegraphics[width=0.95\textwidth]{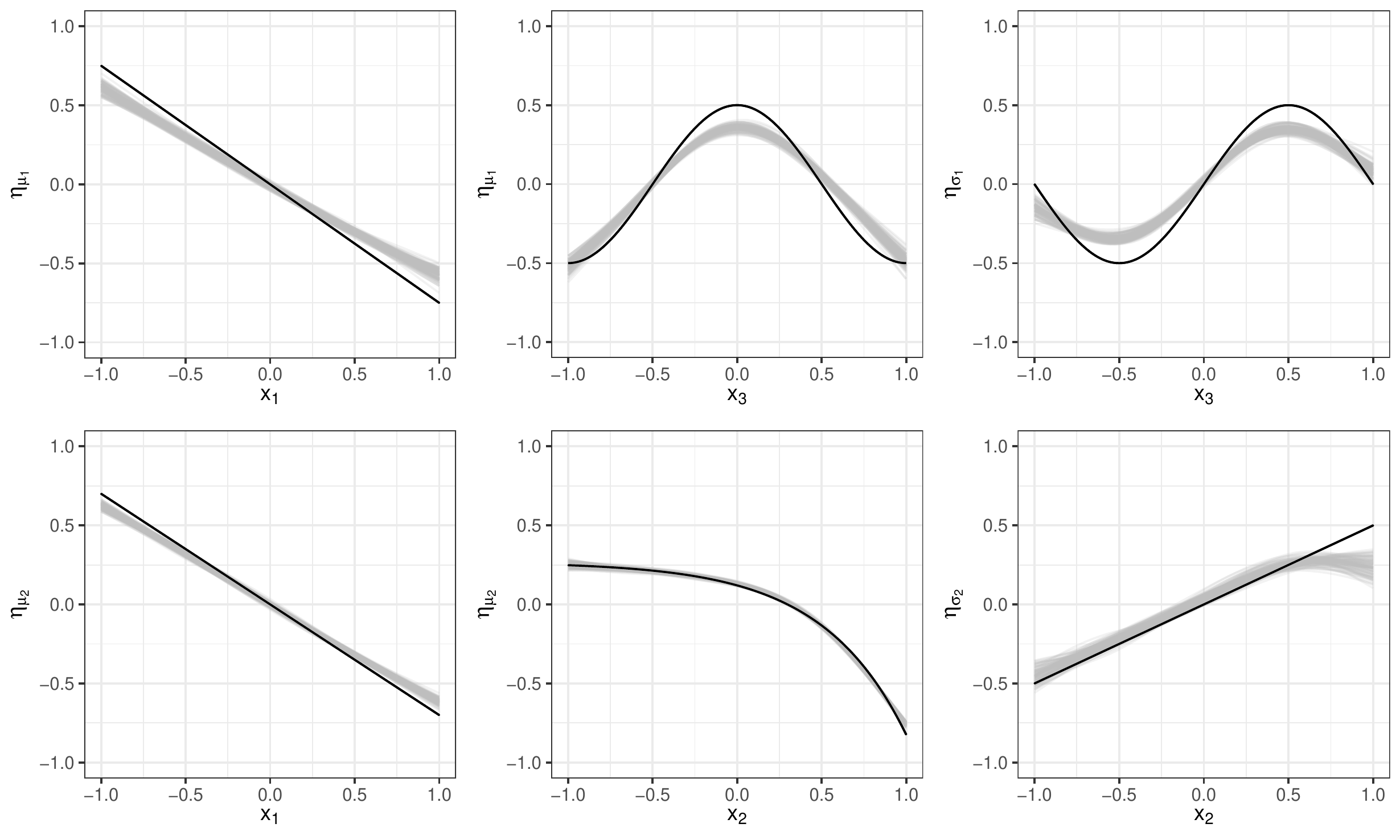} 
\caption{Effect estimates on the marginal models for the Gumbel copula in the high-dimensional setting. The first and the second row of plots show the informative covariate effects on the parameters of the first and the second marginal distribution (i.e.~Log-Normal and Log-Logistic distribution), respectively. In each plot the gray lines represent the estimates of the 100 runs while the black line displays the original effect.}
\label{fig:HighDimMargEst}
\end{figure}

Figure \ref{fig:HighDimMargEst} presents the estimates of the informative covariates on the marginal distribution parameters of the Gumbel copula in the high-dimensional setup. We again only discuss the marginal plots of only one copula as the results do not differ between Gauss, Clayton and Gumbel. The respective plots of the Gaussian and Clayton copula can be found in the Appendix (A.3.1). 
In general, the estimates show the true effect structure for each of the marginal model parameters. In comparison to the low-dimensional setting, the shrinkage effect towards zero is stronger. In general, the non-informative covariate effects again gather around zero as can be seen in the Appendix (A.3.2). 
\begin{figure}[htbp]
\includegraphics[width=0.95\textwidth]{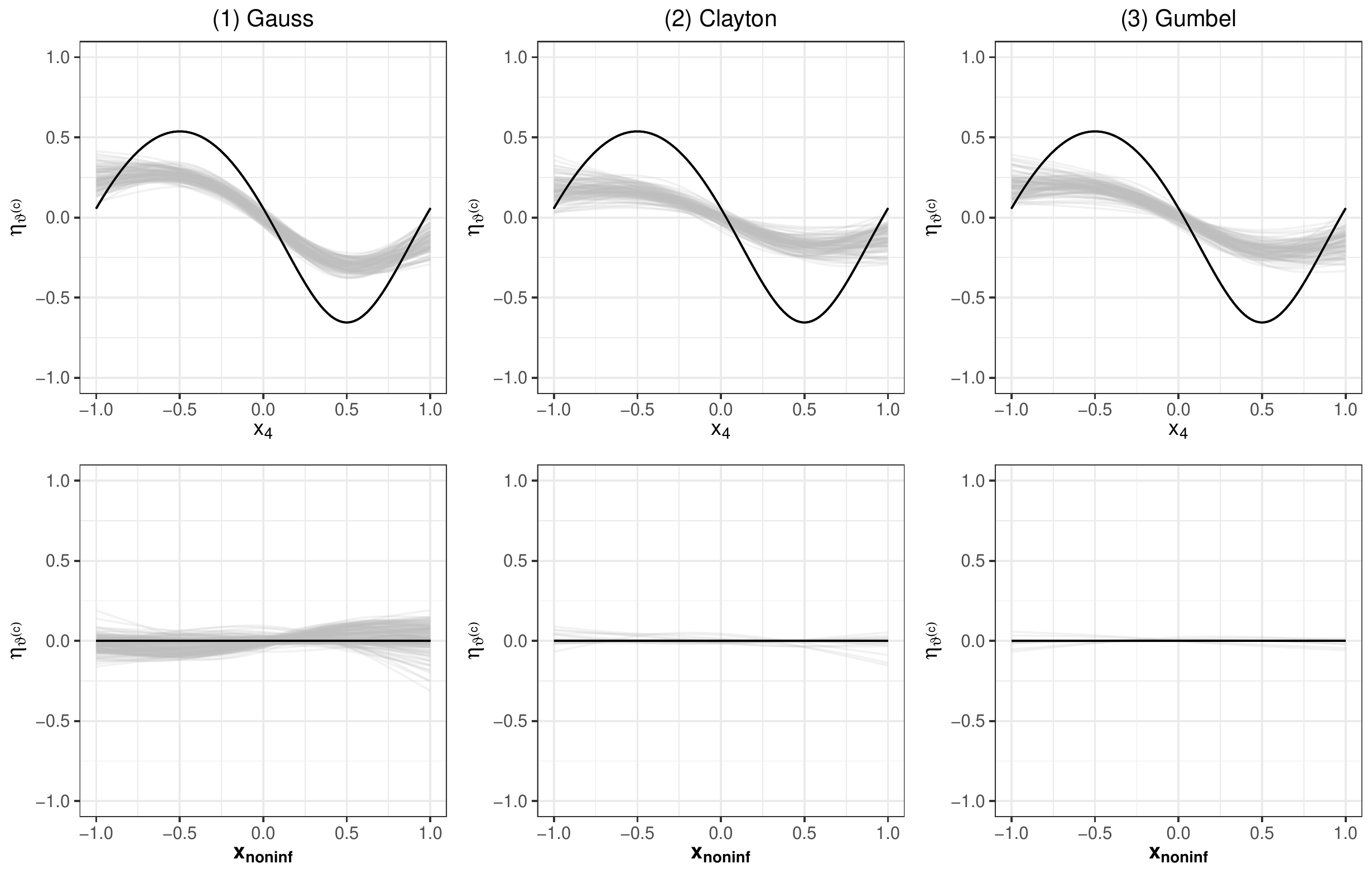} 
\caption{Dependence parameter fits for the Gaussian, Clayton and Gumbel copula in the high-dimensional setting. The first and second row of plots shows the estimates of the informative and the non-informative covariate(s) on the copula parameter, respectively. In each plot the gray lines represent the estimates of the 100 runs while the black line displays the original effect.}
\label{fig:HighDimCopEst}
\end{figure}
For the dependence parameters the covariate effects are given in Figure \ref{fig:HighDimCopEst} for the three copula functions in the high-dimensional setting. Similar to the low-dimensional setting the impact of the informative variable is more shrunk and has a higher variance in comparison to the marginal estimates. Moreover, in case of the Gaussian copula model the effects of the non-informative variables on the dependence parameter tend to be more pronounced then for Clayton and Gumbel.

\subsubsection*{Variable selection performance}
All informative variables are selected in every simulation run for all three copula functions, leading to correct selection rates of $100 \%$. Table \ref{tab:HighDimSelAcc} shows the average selection rates of the non-informative variables in the high-dimensional setting. For the three marginal models $\mu_{1}$, $\sigma_{1}$, $\sigma_{2}$ and the copula dependence model $\vartheta^{(c)}$ the average selection rates are sound for each of the three copula functions. Only in the  $\mu_{2}$ model several non-informative covariates are integrated with an average amount of 67.18 (sd = 12.06), 103.39 (sd = 13.79) and 70.29 (sd = 8.48) selected non-informative variables for Gauss, Clayton and Gumbel, respectively.
\begin{table}[htbp]
\caption{Selection rate of non-informative variables averaged over the 100 simulation runs for the different model parameters and copula functions in the high-dimensional setting.}
\centering
\begin{tabular}{ |l|l|l|l| }
\hline
Model Parameter & Gauss &  Clayton  & Gumbel\\
\hline
$\mu_{1}$ & 0.00\%  & 00.01\% &  00.68\%  \\
\hline
$\mu_{2}$ & 06.73\% & 10.36\% & 07.04\%  \\
\hline
$\sigma_{1}$ & 00.00\% & 00.07\% & 00.15\%  \\
\hline
$\sigma_{2}$ & 00.11\%  & 00.30\% & 00.46\% \\
\hline
$\vartheta^{(c)}$ & 0.18\% & 0.02\% & 0.01\%  \\
\hline
\end{tabular}

\label{tab:HighDimSelAcc}
\end{table}

\subsubsection*{Estimation with independent responses}
For the independent setup the three copula functions do not select any covariate in their dependence parameter model. This result suggests that boosting indeed works well for independent responses in the high-dimensional case. 

\subsubsection*{Predictive risk}
\begin{figure}[htbp]
\includegraphics[width=0.95\textwidth]{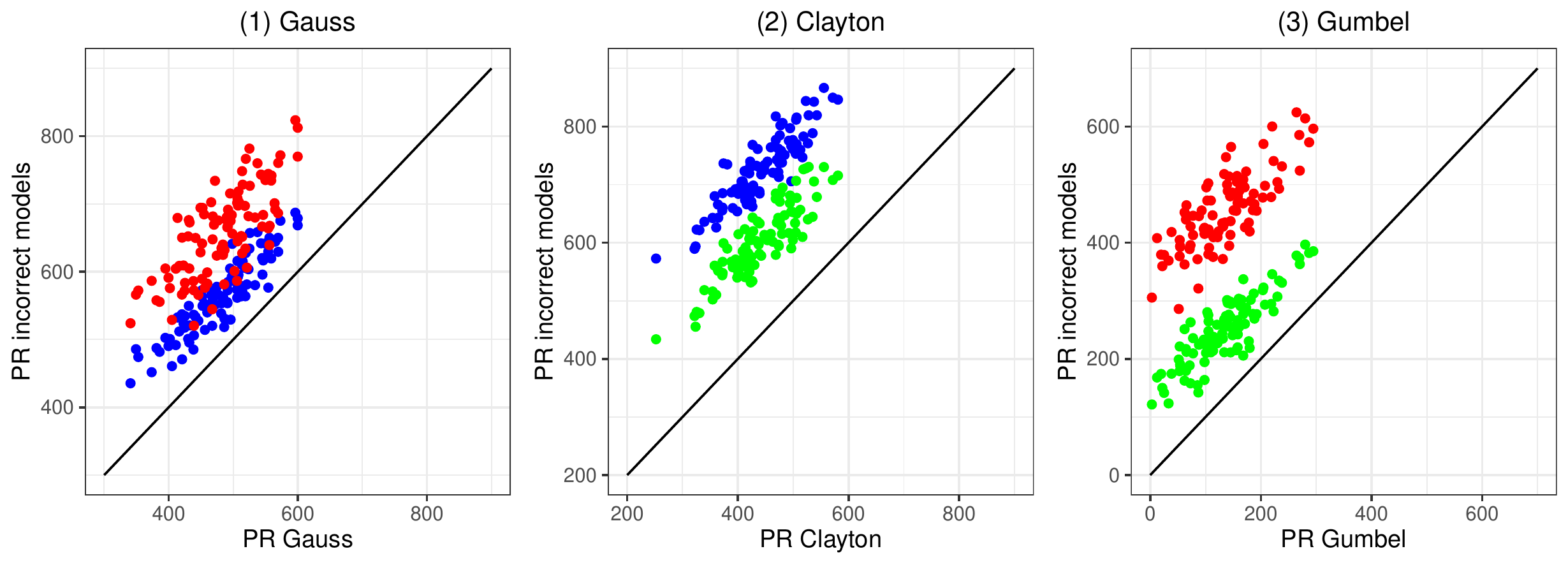}
\caption{Predictive risk values of the correct model (x-axis) versus the incorrect models (y-axis) in the high-dimensional data case. For (a), (b) and (c) the Gaussian, the Clayton and the Gumbel copula is the correct model, respectively. The red points mark the comparison to the incorrect Clayton copula, the blue points to the incorrect Gumbel copula and the green points to the incorrect Gaussian copula.}
\label{fig:HighDimPR}
\end{figure}
Similar to the low-dimensional setup, the predictive risk also performs accurately in the high-dimensional data case. Figure \ref{fig:HighDimPR} shows the values of the predictive risk for the correct copula model versus the values of the predictive risk of the misspecified models. All points are placed above the diagonal line, suggesting that the predictive risk is a helpful tool to discriminate between the true and incorrect copula functions in the high-dimensional setup.

%% file: sec5.tex
\section{Analysis of Fetal Ultrasound Data}
\label{sec:app}

The weight of a fetus is an important factor in clinical obstetrics and gynecology \citep[][]{Bar1997}. Both a very low and a very high weight are associated with increased risks of adverse events during labor. For instance, fetal macrosomia is related to various perinatal complications with respect to the child like shoulder dystocia, brachial plexus injury and even fetal death \citep[][]{BouAleSal2003}. For the mother, problems like protracted labor and postpartum hemorrhage among others are associated with fetal macrosomia \citep{Fer2000}. 
Fetal ultrasound measurements play an important role in the determination of the fetal weight and thus are of major importance for delivery and labor management \citep[][]{Dud2005}. In the current literature and clinical practice, the length of the newborn is not part of the prediction scheme -- but could also lead to a better decision making during the labour process. 
The length and the weight of a fetus are likely to interrelate, which stresses the relevance to model and predict these two responses together. 
This is particularly important, when we aim at evaluating the joint probability that certain crucial thresholds are passed or to identify particular cases with disproportional growth, like a large fetus with low weight or macrosomic infants with small length. Particularly interesting could be also the identification of predictor variables that influence the association between weight and length of the fetus.
Model-based boosting was applied several times in the development of prediction formulas for fetal weight, due to its advantages in prediction situations and its intrinsic variable selection mechanism \citep[][]{SchMarSie2008, FasBecGoe2012, FasDamRaa2016}, but we are the first to consider fetal weight jointly with its height. 
In the following, we analyse the birth length (supine length) and weight in a birth cohort data set from the University Hospital Erlangen by means of boosted copula distributional regression models.
First, the data cohort is thoroughly described. Subsequently, the choices of the marginals and the copula function are outlined and some details on the interpretation of the most appropriate model and the intrinsic variable selection are provided. Finally, we compare the predictive performance of this model to univariate models, i.e.~models that fit the birth length and birth weight independently, and evaluate the joint probabilities to cross two important thresholds. 

\subsection{Birth cohort data}
Previous analyses investigate different facets of sonographic fetal weight prediction by means of the Erlangen birth cohort data \citep[][]{FasDamRaa2015, FasRaaHei2016, FasDamRaa2016}. 
The data set we use for our illustrative analysis was collected at the Department of Obstetrics and Gynecology of the Erlangen University Hospital and consists of $n=$6103 pregnancy observations during 2008-2016.
Since this was a secondary analysis of routine data, no ethics vote was required for this study. Publication or distribution of the original data, however, is not allowed following legal restrictions on protection of personal data.
For our analysis, only singleton pregnancies are included with cephalic presentation and an ultrasound examination with complete biometric parameters that are measured within a time interval of 0–-7 days before delivery. Furthermore, observations with any chromosomal or structural anomalies and intrauterine fetal deaths are excluded. At the Erlangen University Hospital a sonographic estimation is routinely performed in all patients that register for delivery. The sonographic variables we include in our analysis are abdominal anteroposterior diameter, abdominal transverse diameter, abdominal circumference, biparietal diameter, occipitofrontal diameter, head circumference and femur length and the interactions between these covariates. Additionally, we consider the clinical variables weight, height and body-mass-index (BMI) of the mother, gestational diabetes (that is if the mother suffered from gestational diabetes during pregnancy), gravida (i.e.~number of pregnancies of the mother), para (i.e.~number of children of the mother), the sex of the fetus and the gestational age. This leads to a total amount of 36 covariates. The response variables birth length and birth weight are determined within 1 hour after delivery by the nursing staff and are measured in centimeters and kilograms, respectively.

\subsection{Model building}
First of all, we are concerned with the choice of the marginal distributions and the copula function. Following the section 'Model building', we randomly assign the data to a training ($n_{train} = 4103$) and a test ($n_{test} = 2000$) data set and choose the optimal marginal distributions and the copula function by means of the predictive risk. For both response variables we compare the performance of the Gamma, Inverse Gaussian, Log-Normal, Log-Logistic and Weibull distribution, as birth length and birth weight are continuous on $\dsR^{+}$.  
Estimation is performed with the noncyclic version of the model-based boosting algorithm implemented in the \texttt{gamboostLSS} package. We set the step length to $s = 0.01$ and determine the optimal stopping iteration $m_{\mbox{\scriptsize{stop}}}$ via 10-fold cross-validation. Moreover, to ensure similar effective step-lengths among outcomes and parameter dimensions, we apply a gradient stabilization \citep{HofMay2016}. In all model fits we use cubic P-splines with 20 equidistant knots, a second-order difference penalty and 4 degrees of freedom as base learners for all continuous variables and include the categorical variables sex of the fetus and gestational diabetes via linear base learners. 
Table \ref{tab:PRMarg} shows the optimal stopping iteration $m_{\mbox{\scriptsize{stop}}}$ and the corresponding predictive risk of the five marginal distributions for the two responses. For both birth length and birth weight the Log-Logistic distribution suits best with regard to the predictive risk. 
\begin{table}[H]
\caption{Predictive risk values and optimal stopping iterations for the Gamma, Inverse Gaussian, Log-Normal, Log-Logistic and Weibull distribution for birth length and birth weight. The Log-Logistic distribution yields the lowest predictive risk values for both birth length and birth weight (in green).}
\centering
\begin{tabular}{ |l|l|l|l|l| }
\hline
Distribution & \multicolumn{2}{ c| }{Birth length} & \multicolumn{2}{ c| }{Birth weight}\\
& Predictive risk & $m_{\mbox{\scriptsize{stop}}}$ & Predictive risk & $m_{\mbox{\scriptsize{stop}}}$ \\
\hline
Gamma & 4133.251 & 1009 & 376.948 & 3088  \\
Inverse Gaussian & 4697.663 & 426 & 393.370 & 2463 \\
Log-Logistic & \textcolor{green}{4106.650} & 1291 & \textcolor{green}{348.827} & 2957 \\
Log-Normal & 4137.856 & 424 & 384.000 & 2085  \\
Weibull & 4320.790 & 530 & 523.475 & 1716 \\
\hline
\end{tabular}

\label{tab:PRMarg}
\end{table}
Next, we decide on the most appropriate copula function by fitting the Gaussian, the Clayton and the Gumbel copula with the selected Log-Logistic marginal distributions to the data and comparing the corresponding predictive risks. The optimal $m_{\mbox{\scriptsize{stop}}}$ values are again obtained via 10 fold cross-validation. The $m_{\mbox{\scriptsize{stop}}}$ and predictive risk values are provided in Table \ref{tab:PRCopula}, indicating that the Gaussian copula is the most appropriate choice for this data situation. 
\begin{table}[H]
\centering
\caption{Predictive risk values and optimal stopping iterations for the Gaussian, the Clayton and the Gumbel copula function with Log-Logistic marginal distributions for both birth length and birth weight. The Gaussian copula function yields the lowest predictive risk value (in green).}
\begin{tabular}{ |l|l|l| }
\hline
Copula & PR & $m_{\mbox{\scriptsize{stop}}}$\\
\hline
Clayton & 4271.954 & 5381 \\
Gauss & \textcolor{green}{4157.042} & 6278 \\
Gumbel & 4162.701 & 6633 \\
\hline
\end{tabular}

\label{tab:PRCopula}
\end{table}

\subsection{Results and  selected variables}

\subsubsection*{Marginals}
For the $\mu$ models of the Log-Logistic distributions for birth length and birth weight the boosting algorithm selects $0.92 \%$ and $0.81 \%$ of the covariates, respectively. Lower selection rates are achieved for the $\sigma$ models with $0.44 \%$ for the birth length and $0.39 \%$ for the birth weight. 
\begin{figure}[htbp]
\includegraphics[width=0.98\textwidth]{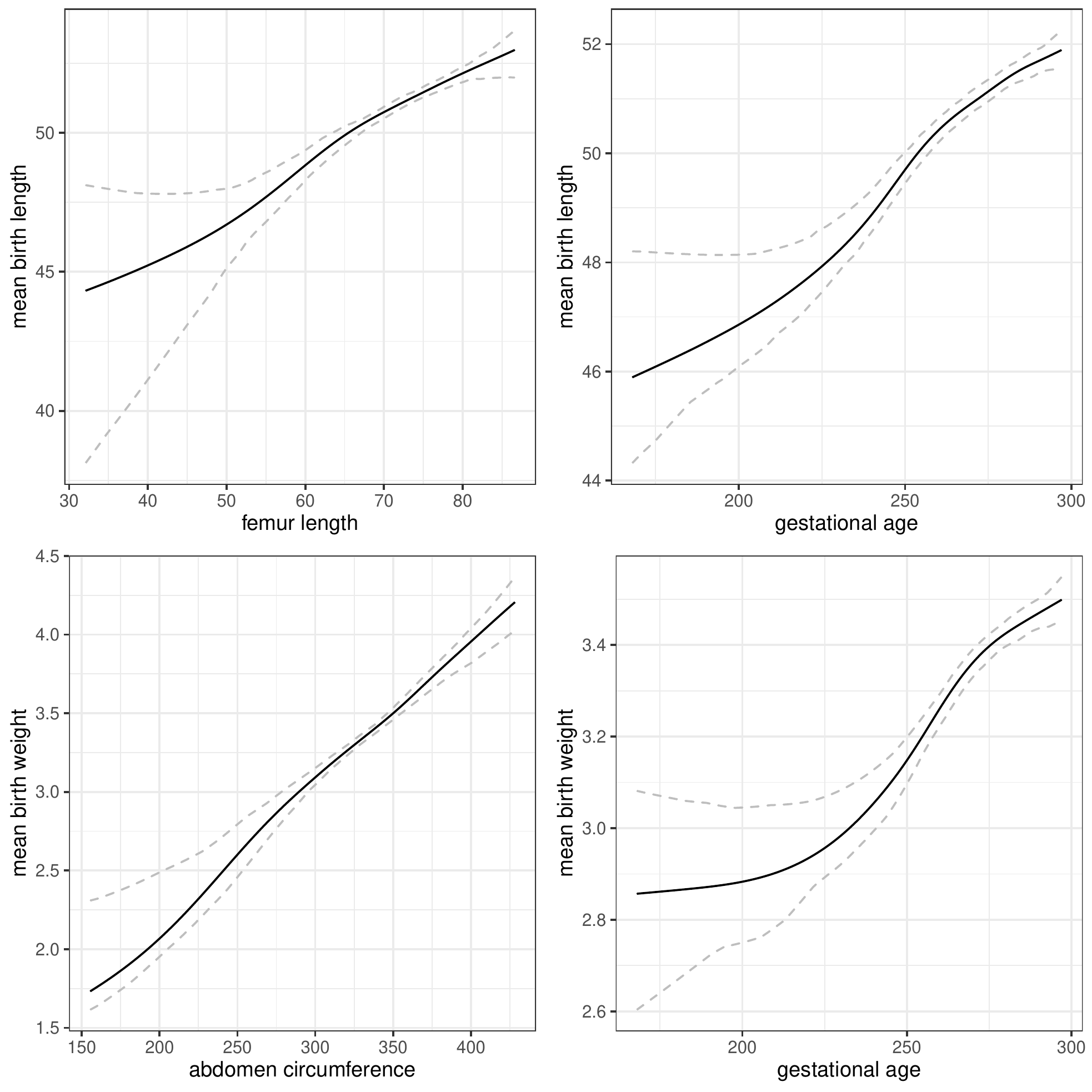} 
\caption{Effects of femur length and gestational age on the mean birth length in the first row of plots and effects of abdomen circumference and gestational age on the mean birth weight in the second row of plots (solid line). All other covariates are set to the average covariate values in case of continuous variables and to the reference group in case of binary variables. The dashed lines represent the 95\% confidence bands estimated from 100 bootstrap samples.}
\label{fig:BirthDataMarg}
\end{figure}
These results correspond to the findings of the simulation study of the section 'Simulations', where the $\mu$ models tent to select also many non-informative covariates (false positives). For all parameter models mostly sonographic covariates and gestational age are selected. 
The plots of Figure \ref{fig:BirthDataMarg} show the effects of femur length and gestational age on the mean birth length and abdominal circumference and gestational age on the mean of birth weight as defined in equation~\eqref{eq:meanloglog}. For each plot all other covariates are set to the reference category in case of the binary variables and set to the average covariate values of the whole data set in case of continuous variables. The mean birth length and the mean birth weight increase with higher covariate values for all presented plots.
The effects of the sonographic covariates femur length and abdomen circumference are almost linear. These results are coherent with the findings of previous univariate sonographic birth weight estimations \citep[][]{FasDamRaa2016, FasBecGoe2012}. 

\subsubsection*{Dependence parameter and Kendall's tau}
The boosting algorithm selects $25\%$ of the covariates into the dependence parameter model. Mostly sonographic covariates determine the relation between birth length and birth weight.
\begin{figure}[htbp]
\includegraphics[width=0.98\textwidth]{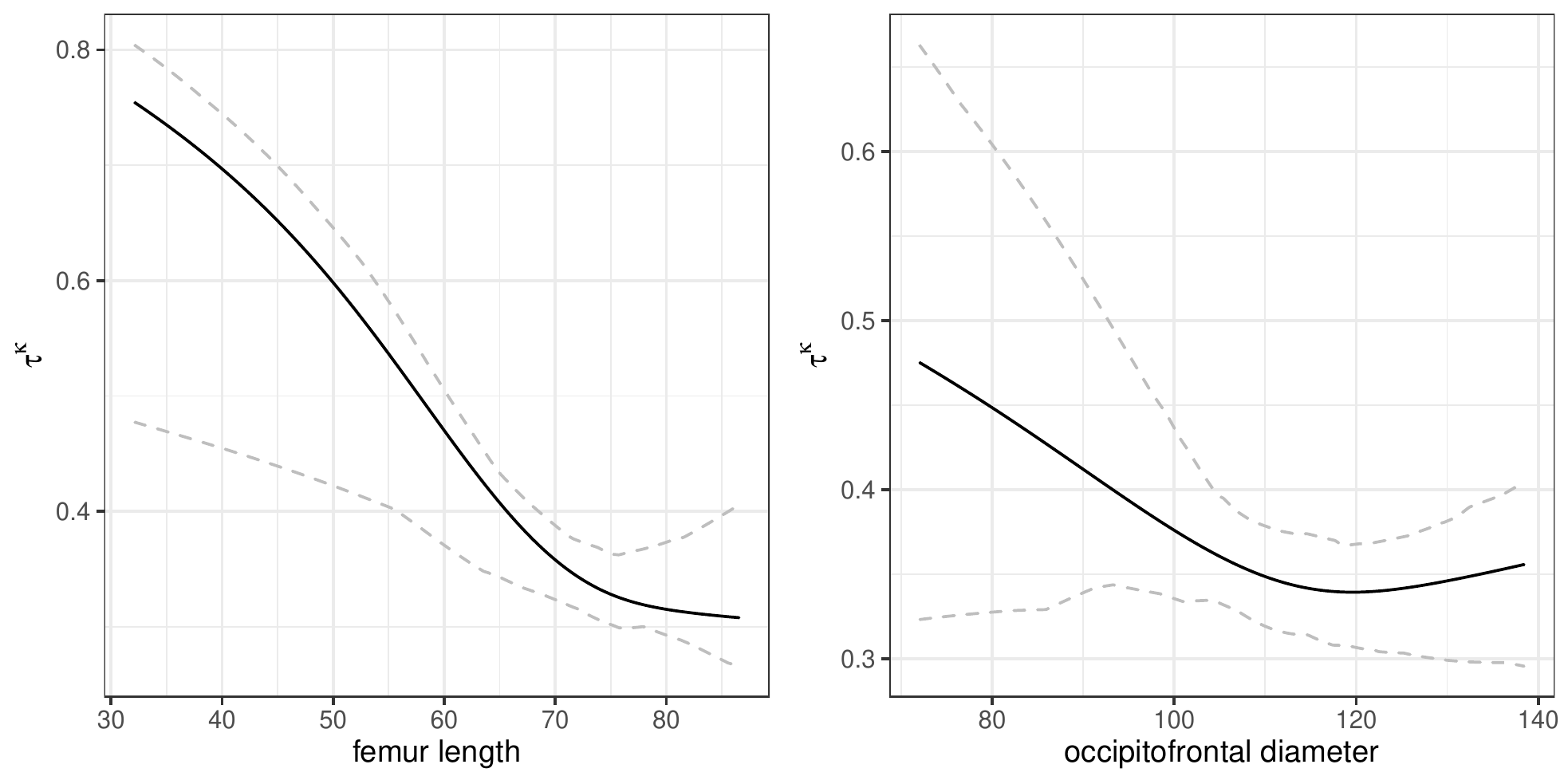}
\caption{Effects of femur length and occipitofrontal diameter on kendall's tau rank correlation coefficient $\tau^{\kappa}$ between birth length and weight (solid line). All other covariates are set to the average covariate values in case of continuous variables and to the reference group in case of binary variables. The dashed lines represent the 95\% confidence bands estimated from 100 bootstrap samples.}
\label{fig:BirthDataRho}
\end{figure}
For both femur length and occipitofrontal diameter kendall's tau rank correlation coefficient $\tau^{\kappa}$ decreases for higher covariate values as presented in Figure \ref{fig:BirthDataRho}. 
These findings seem plausible, as an increase of femur length possibly has a stronger impact on the birth length than on the birth weight. 
For the occipitofrontal diameter we assume that a larger head of a fetus rather effects the birth weight than the birth length, leading to a weaker association.

\subsection{Comparison to univariate models}
We want to compare the probabilistic forecasting performance of the Gaussian copula model with univariate distributional regression models for the birth length and the birth weight by fitting the two univariates responses independently from each other. For the univariate models we specify a Log-Logistic distribution for both response variables and use the \texttt{gamboostLSS} package for estimation. 
In order to compare the bivariate model with the univariate counterparts, we use multivariate proper scoring rules. 
These are multivariate extensions of univariate scoring rules assigning a numerical score to a combination of a future observation and its predictive probabilistic distribution \citep[][]{GneRaf2007}.  As such, scoring rules can also be applied to compare competing models. In particular, we focus on two scoring rules, namely the logarithmic score and the energy score. 
The logarithmic score is defined as the negative predictive log likelihood of a future event and the forecast probability distribution and thus lays a focus on the predictive density \citep[][]{GneRaf2007, JorKruLer2019}.
The energy score is a multivariate generalization of the univariate continuous ranked probability score \citep[][]{GneStaGri2008}, which is defined in terms of the predictive cumulative distribution function \citep[][]{GneRaf2007}. We calculate the averaged score values of the 2000 observations in the test data for the logarithmic score and the energy score for the Gaussian copula model and the independent univariate models. In case of the energy score we use the implementation of the R package \texttt{scoringRules} \citep{JorKruLer2019}. 
Both scores favour the Gaussian copula model over the independent univariate models for our data set. The resulting values for the logarithmic score are 2.079 and 2.228 for the Gaussian copula and independent models, respectively. For the energy score the  former has an average score of 1.085, while the latter yields an average score of 1.087. Note that for both scores lower values indicate a better probabilistic forecasting performance \citep{JorKruLer2019}.

\subsection{Joint probabilities and thresholding}
\begin{figure}[htbp]
\includegraphics[width=0.95\textwidth]{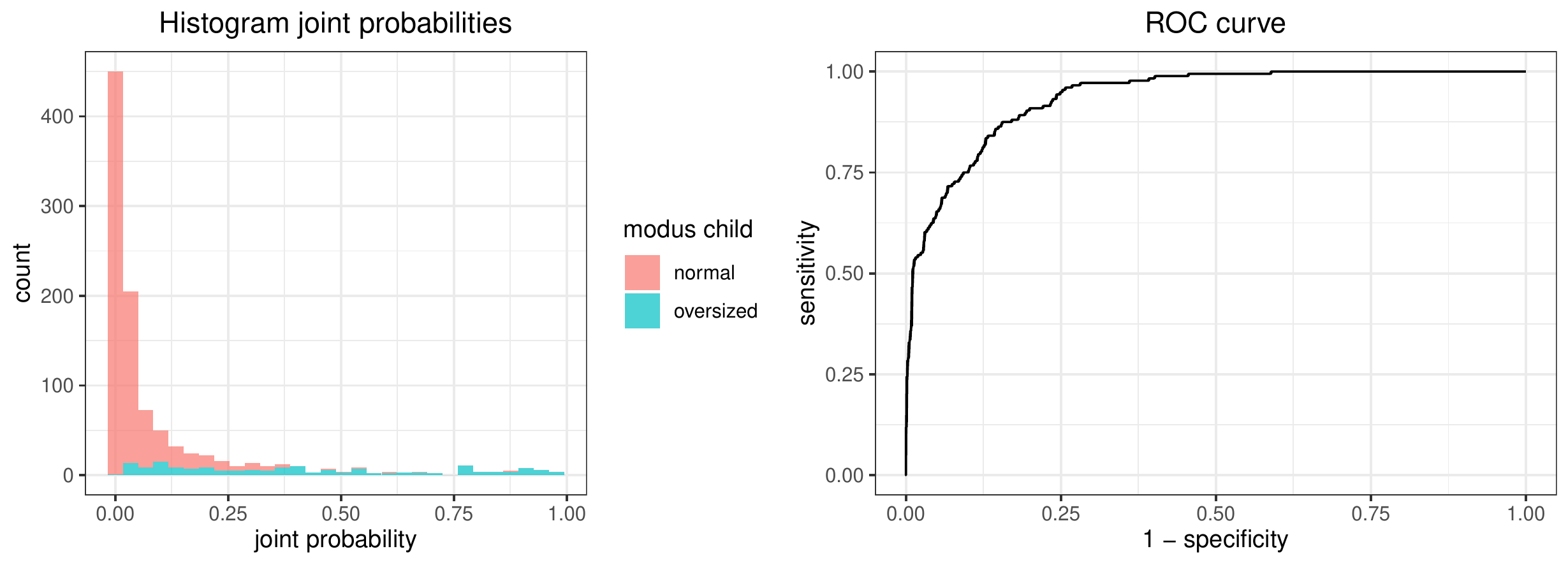} 
\caption{Joint probabilities from the copula model. Left: histogram of  joint probabilities separated by oversized children (blue) and children that meet standard values for length and weight (red) as defined in the main text. Right: ROC over the joint probabilities.}
\label{fig:JointProb}
\end{figure}

Finally, we evaluate the predictive joint probabilities of the fetuses from the test sample to exceed a birth length of 51 cm and a birth weight of 4 kilogram for the Gaussian copula model. 
A birth weight of $ > 4$ kilogram is defined as fetal macrosomia \citep[][]{FasDamRaa2015}. Following a study on international standards on newborns' length  \citep[][]{VilCheOhu2014} the 90-th centile of the birth length is given by approximately $51$ cm for fetuses within the 38th and 41th week of gestational age. These two thresholds are interesting to investigate as it is for example more likely for mothers with large babies to experience a cesarean section \citep[][]{BouAleSal2003}. The Erlangen birth data set confirms this finding by a sectio rate that is approximately 1.42 times higher for newborns that exceed these thresholds versus newborns that meet standard values, i.e.~birth length $\in [47  ,51]\text{cm}$ and birth weight $ \in [2.65 , 4 ]\text{kg}$, where 47 cm and 2.65 kg are approximately the 10-th centiles for birth length and birth weight for fetuses within the 38th and 41th week of gestational age, respectively \citep[][]{VilCheOhu2014}. 
Figure \ref{fig:JointProb} (left) shows the histogram of the joint probabilities separated by oversized children (blue) and children that meet standard values for length and weight (red) in the test data set. It is obvious that the joint probabilities to exceed both thresholds tend to be higher for oversized children. Thus, we consider the receiver operating characteristic curve (ROC, right) to evaluate the classification performance of the joint probabilities. With an area under the curve (AUC) of 0.937 the joint probability seems to be a valuable indicator to identify oversized children. 

%% file: sec6.tex
\section{Conclusion}\label{sec:conclusion}

We have introduced an extremely flexible framework to fit conditional copula regression models via a model-based boosting algorithm. 
Boosting proves to be a valuable fitting procedure in many applied data analysis settings, due to its fully data-driven mechanism to select variables and predictor effects, its feasibility in high-dimensional data settings and its good performance in prediction setups. In extensive simulation studies we demonstrate the sound performance of boosted conditional copula regression models in low and high dimensional settings. Moreover, an application on fetal ultrasound data exemplarily illustrates the potential of boosted copula regression in medical research. Boosted copula regression could be of immense interest in many other applications of biomedical research and beyond. In many clinical trials or experiments, one typically evaluates a potential intervention effect on various outcome variables (primary or secondary endpoints), which are typically analysed separately. This is not necessarily problematic \citep{Zel1962}, however, it is hence also not possible to detect effects on the complex association of these endpoints. \\
We have focused on three popular bivariate one-parameter copula functions, i.e.~the Clayton, Gauss and Gumbel copulas, and two continuous marginal distributions driven by our application, i.e.~the Log-Logistic and Log-Normal. 
However, due to the generic and modular nature of boosted copula regression models, our algorithm is readily applicable to other copula and marginal distribution functions. \\
Despite its merits, we notice three major limitations of our approach that frequently occur in the context of boosting regression models in general. First of all, even though statistical boosting incorporates an intrinsic variable selection mechanism, in some situations the algorithm tends to include too many variables. This selection behaviour occurs in particular in low-dimensional data settings, where boosting shows a slow overfitting behavior and a new procedure for enhanced variable selection is proposed that effectively deselects base-learners with minor importance \citep{StrStaKle2021}. Secondly, boosting distributional regression with a unique step length for all parameter submodels might lead to imbalanced updates of predictors in some scenario as \citet{ZhaHepGre2022} outline. In certain situations this can become a problem as some submodels might not be appropriately fitted within a limited number of boosting iterations. As a result the authors propose using adaptive step-lengths in a univariate GAMLSS regression, which proves to be beneficial in Gaussian location and scale models but has up to date not been tested extensively in other GAMLSS family classes. Finally, a further limitation of model-based boosting is its computationally expensive tuning procedure based on cross validation \citep[][]{MayFenHof2012}. Alternatives have been proposed based on stopping the algorithm first a variable from a set of artificial shadow-variables (random probes) is selected instead of the original ones \citep{ThoHepMay2017}. This procedure avoids time-consuming resampling procedures, but leads to a higher memory demand. It will thus be of interest to investigate these approaches in the context of boosted distributional copula regression as well  to potentially overcome or alleviate the respective limitations.\\
A further direction for future research that we aim to explore is the extension to boosted copula regression  models with discrete and continuous-discrete outcomes to leverage Bayesian and frequentist counterparts \citep{RadMarWoj2016,KleKneMar2019,KleKneMarRad2020}. Moreover, it would be interesting to analyze multivariate settings of higher dimension or to investigate enhanced procedures to determine the optimal set of predictor variables like stability selection \citep{MeiBuh2010, ThoMayBis2018} in a boosted copula regression context. 

%% file: append.tex
\newpage
\appendix

\noindent

\section{Further Simulation Results}
\subsection{Converged models}\label{appen:ConMod}

We define an exclusively linear setting with parameter submodels given by
\begin{equation*}\begin{aligned}
    h_{\mu_{1}}(\mu_{1}) &= -1  x_{1} + 0.5 x_{3}
    &h_{\mu_{2}}(\mu_{2}) &= 0.5 - 0.7 x_{1}  + 0.3 x_{2} \\
    h_{\sigma_{1}}(\sigma_{1}) &= -0.7 + 0.7  x_{3}
    &h_{\sigma_{2}}(\sigma_{2})& = 2 + 0.5 x_{2}\\
  h_{\vartheta^{(c)}}(\vartheta^{(c)}) &= 1 + 1 x_{4}.
\end{aligned}\end{equation*}
\begin{figure}[htbp]
\includegraphics[width=0.98\textwidth]{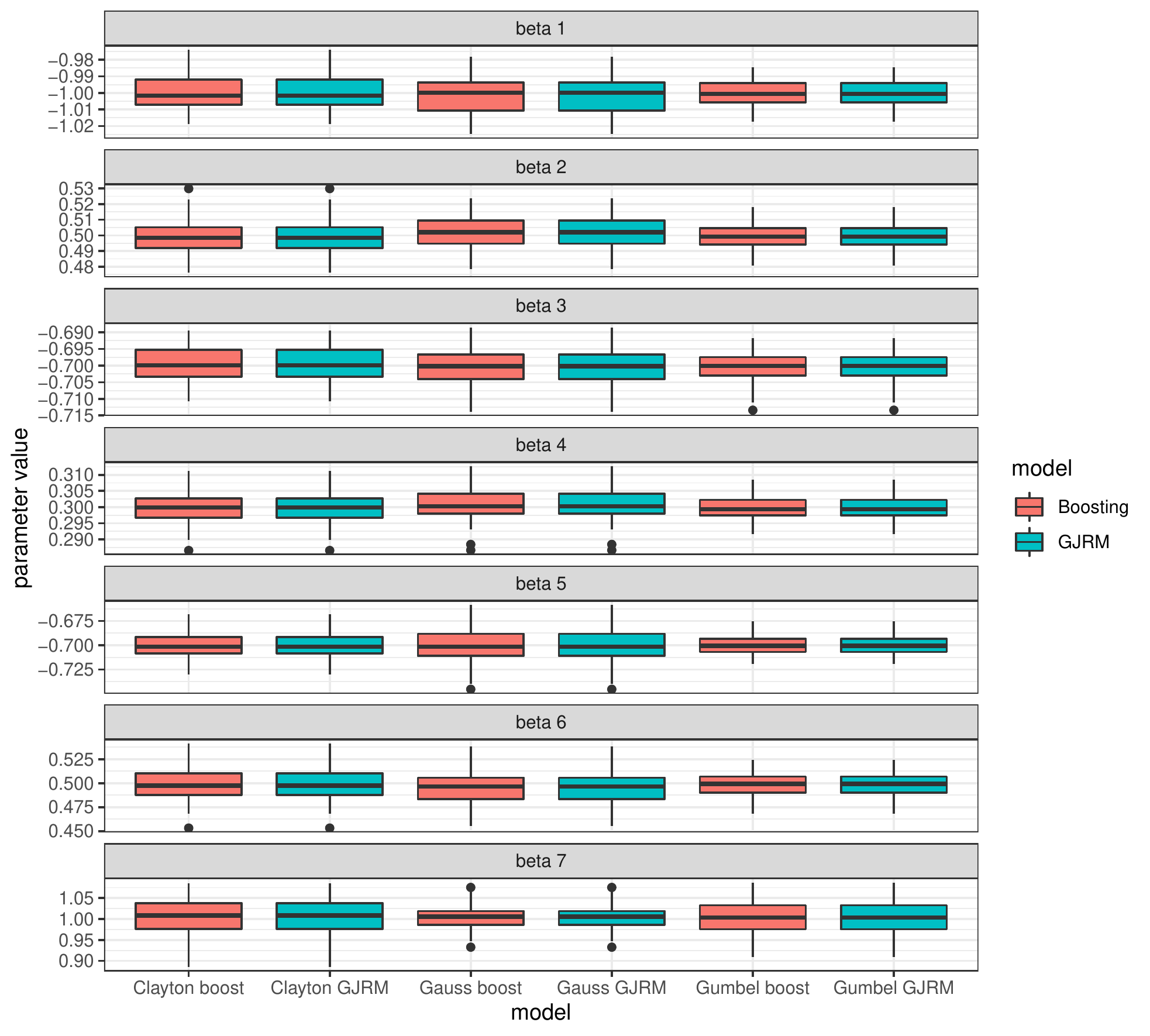}
\caption{Boxplots of the hundred estimated coefficients $\beta_1, \dots, \beta_7$ for the three copulas, Clayton, Gauss and Gumbel, and the two fitting procedures boosting (green) and \texttt{GJRM} (red).}
\label{fig:ConvPlot}
\end{figure}
Hence, the model coefficients are defined as $\beta_1 = -1$, $\beta_2 = 0.5$, $\beta_3 = -0.7$, $\beta_4 = 0.3$, $\beta_5 = 0.7$, $\beta_6 = 0.5$ and $\beta_7 = 1$.\\
Next, we let the boosting algorithm for the Clayton, the Gaussian and the Gumbel copula model in combination with one Log-Normal and one Log-Logistic marginal converge and compare the parameter estimates to their counterparts estimated via the penalized-likelihood based conditional copula regression method implemented in the package \texttt{GJRM}. 
Figure~\ref{fig:ConvPlot} shows the boxplots of the estimated parameters for the different copulas and the two fitting procedures for the 100 simulation runs. The rows of plots represent the estimates of the different parameters $\beta_1, \dots, \beta_7$. 
We find that the boosting estimates indeed converge to the \texttt{GJRM} estimates for all copulas and all coefficients, once we do not apply early stopping. 
The differences in the parameter estimates are on average less than $1e^{-7}$ for all coefficients and all copula models.

\clearpage

\subsection{Further results for the low-dimensional Setting}

\subsubsection{Marginal estimates informative for Clayton and Gumbel}\label{appen:LowDimMargEstInf}

\subsubsection*{Clayton}
\begin{figure}[H]
\includegraphics[width=0.95\textwidth]{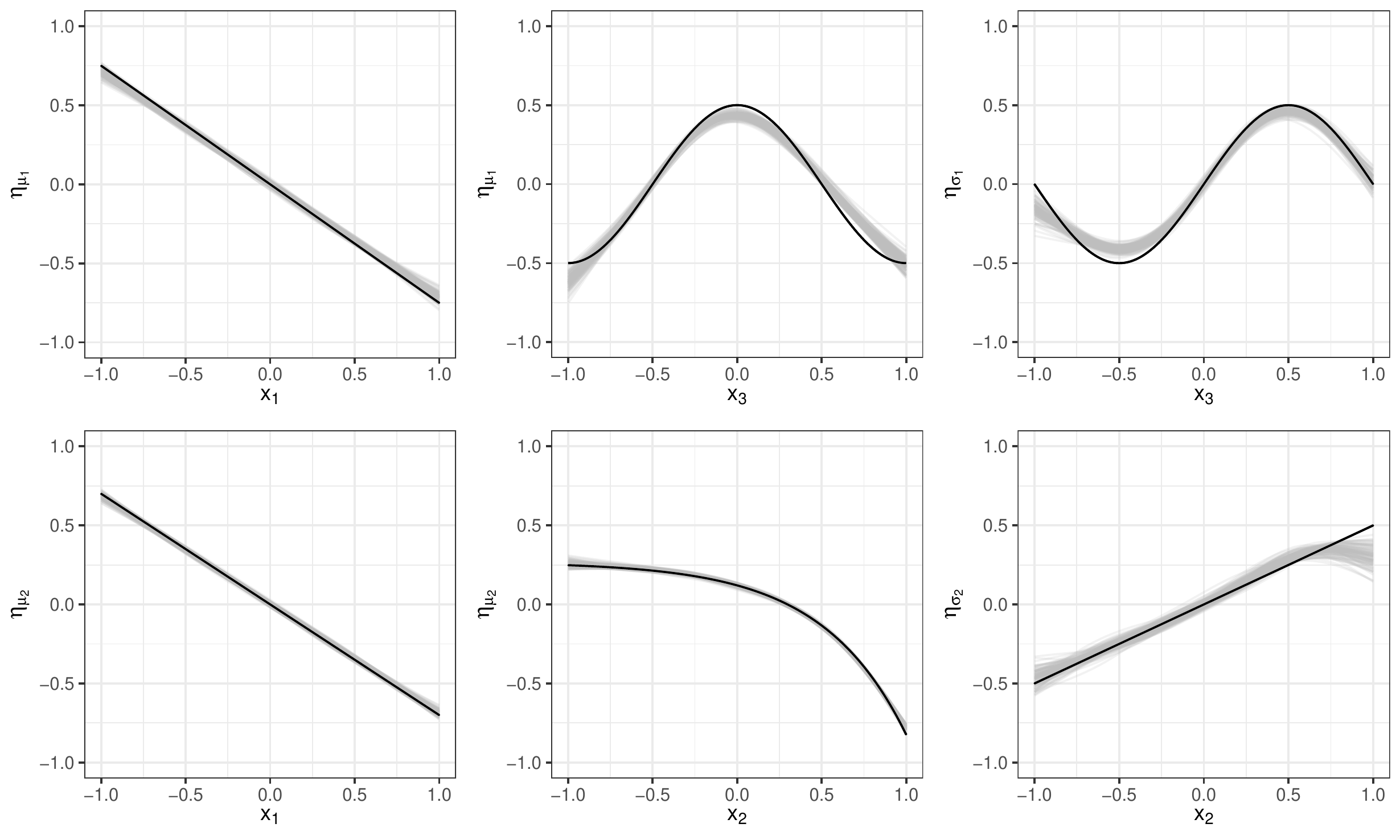}
\caption{Effect estimates of the marginal models for the Clayton copula in the low-dimensional setting. The first and the second row of plots show the effect estimates for informative covariates on the parameters of the first and the second marginal distribution (i.e.~Log-Normal and Log-Logistic distribution), respectively. In each plot the gray lines represent the estimates of the 100 runs while the black line displays the original effect.}
\label{fig:LowDimClay}
\end{figure}

\subsubsection*{Gumbel}
\begin{figure}[H]
\includegraphics[width=0.95\textwidth]{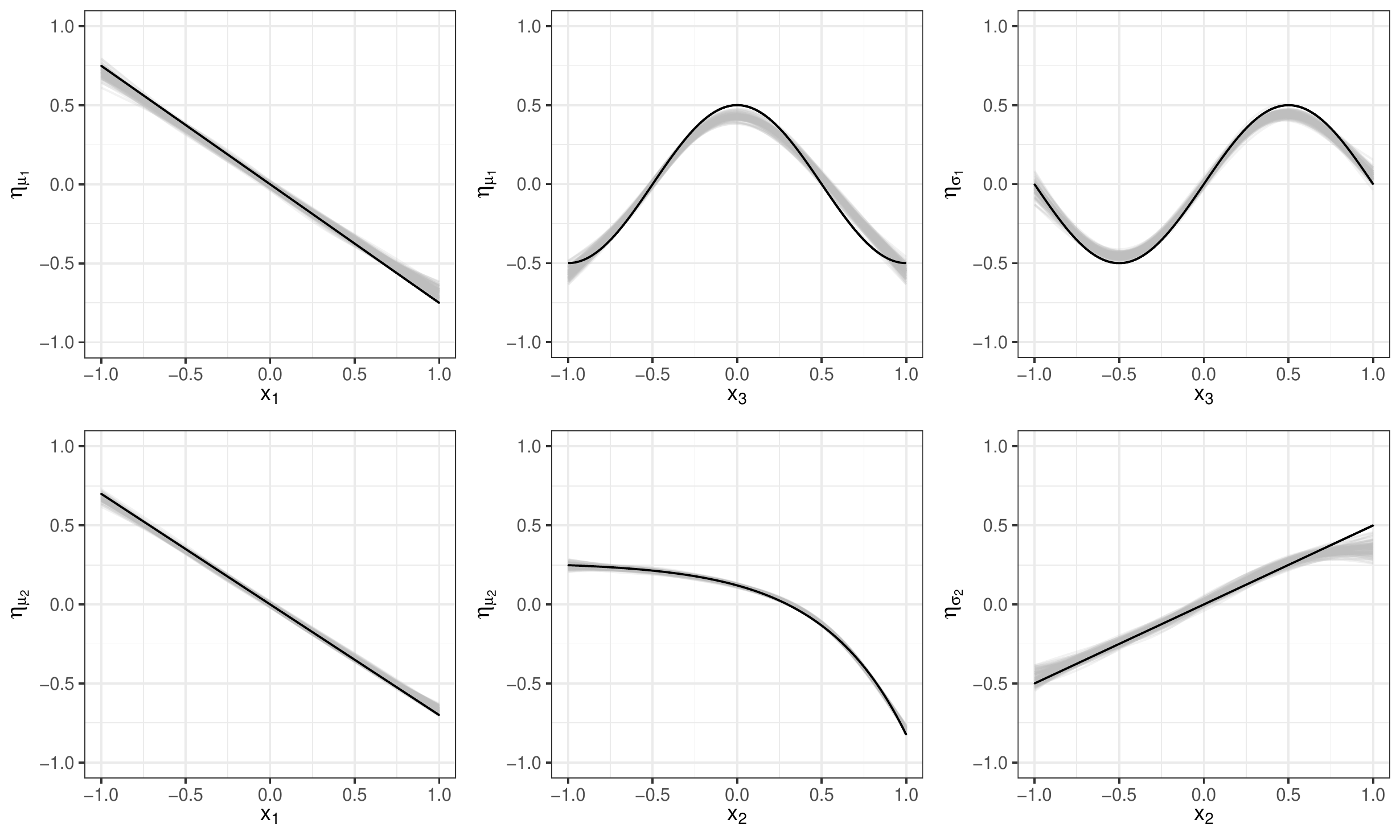}
\caption{Effect estimates of the marginal models for the Gumbel copula in the low-dimensional setting. The first and the second row of plots show the effect estimates for informative covariates on the parameters of the first and the second marginal distribution (i.e.~Log-Normal and Log-Logistic distribution), respectively. In each plot the gray lines represent the estimates of the 100 runs while the black line displays the original effect.}
\label{fig:LowDimGumb}
\end{figure}

\subsubsection{Marginal estimates non-informative for Gauss, Clayton and Gumbel}\label{appen:LowDimMargEstNonInf}

\begin{figure}[H]
\includegraphics[width=0.95\textwidth]{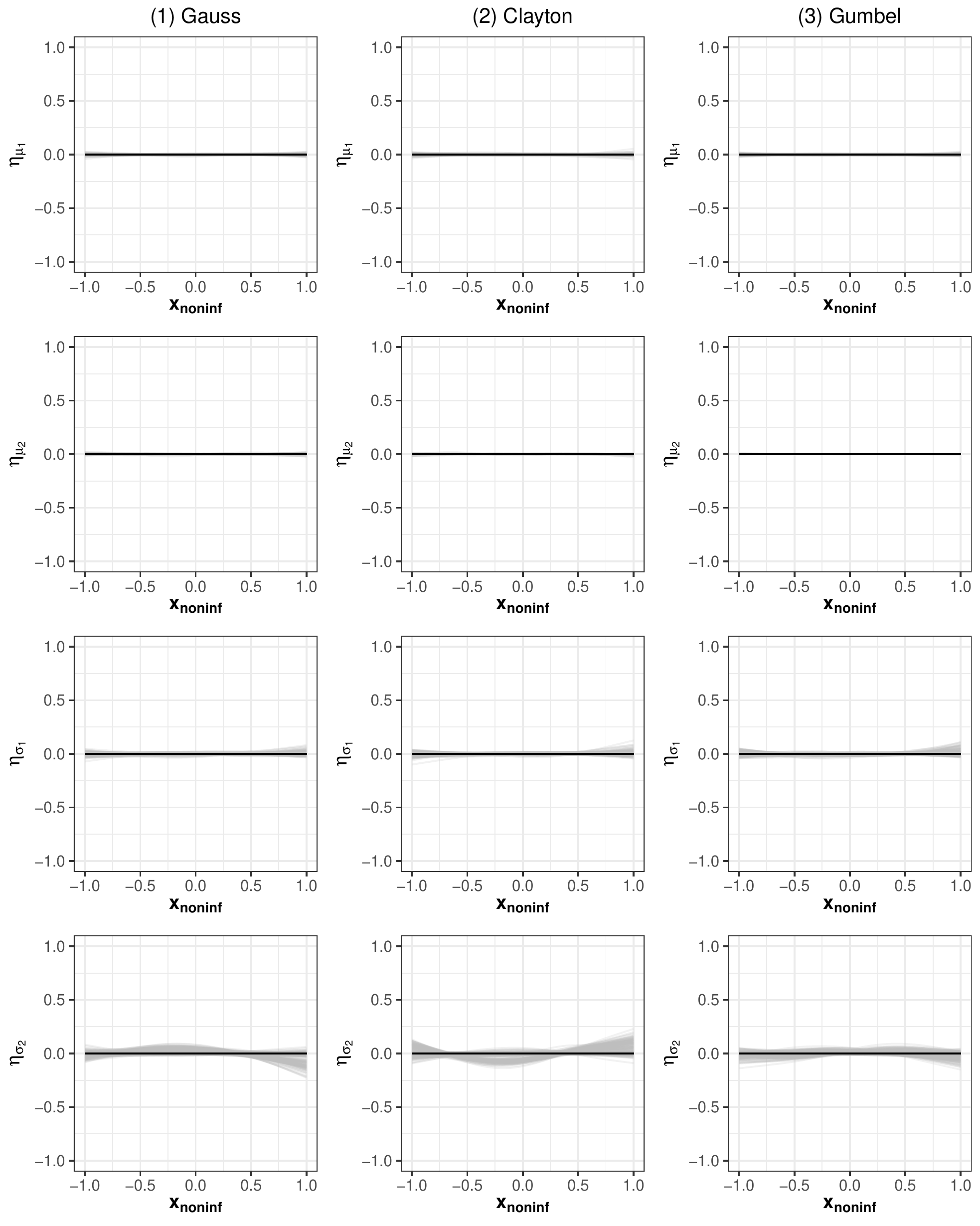}
\caption{Effect estimates of the non-informative covariates on the marginal distribution parameters for the Gaussian, Clayton and Gumbel copula in the low-dimensional setting. In each plot the gray lines represent the estimates of the 100 runs while the black line displays the original effect.}
\label{fig:LowDimNonInf}
\end{figure}

\clearpage

\subsection{Further results for the high-dimensional setting}

\subsubsection{Marginal estimates informative for Clayton and Gauss}\label{appen:HighDimMargEstInf}

\subsubsection*{Clayton}
\begin{figure}[H]
\includegraphics[width=0.98\textwidth]{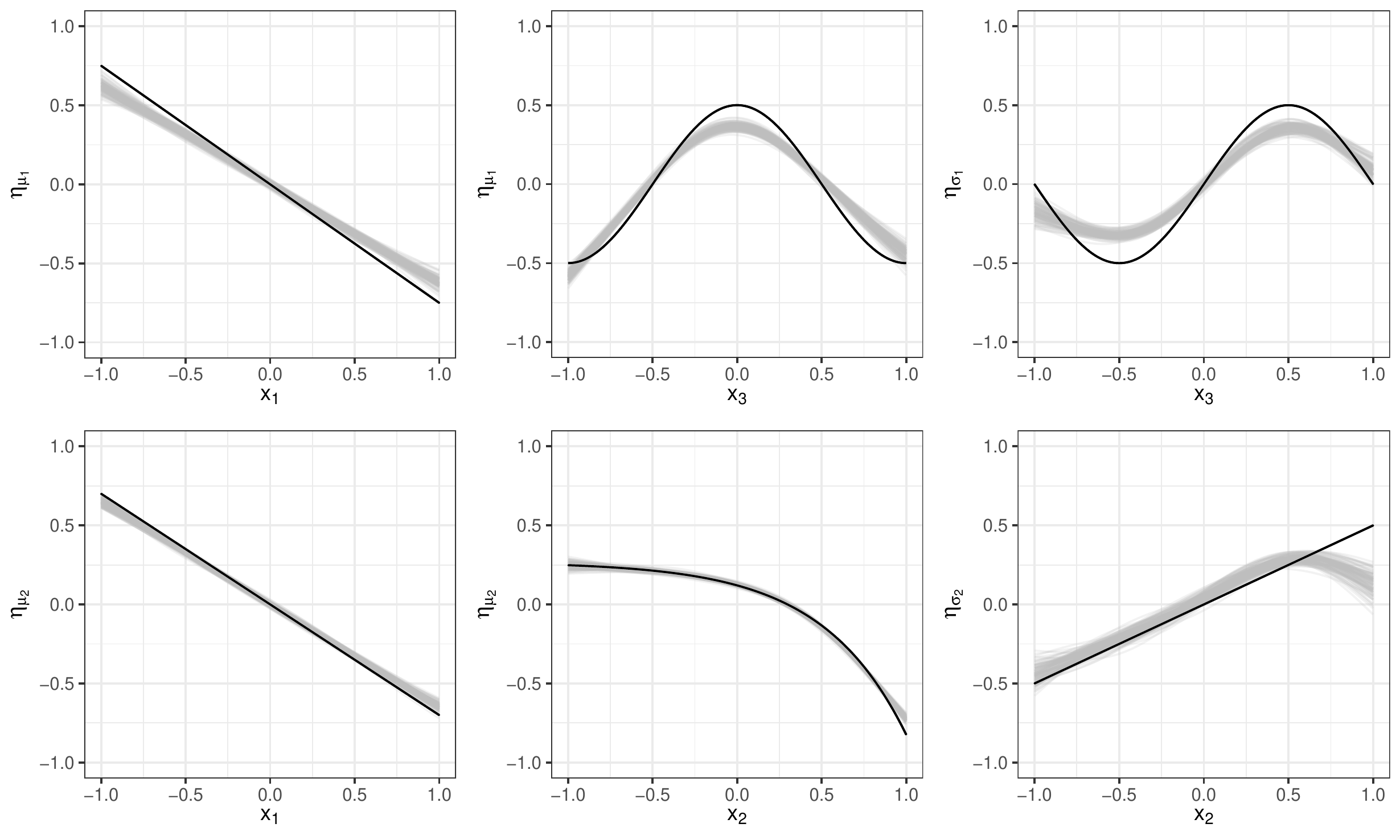}
\caption{Effect estimates on the marginal models for the Clayton copula in the high-dimensional setting. The first and the second row of plots show the informative covariate effects on the parameters of the first and the second marginal distribution (i.e.~Log-Normal and Log-Logistic distribution), respectively. In each plot the gray lines represent the estimates of the 100 runs while the black line displays the original effect.}
\label{fig:HighDimClay}
\end{figure}

\subsubsection*{Gauss}
\begin{figure}[H]
\includegraphics[width=0.98\textwidth]{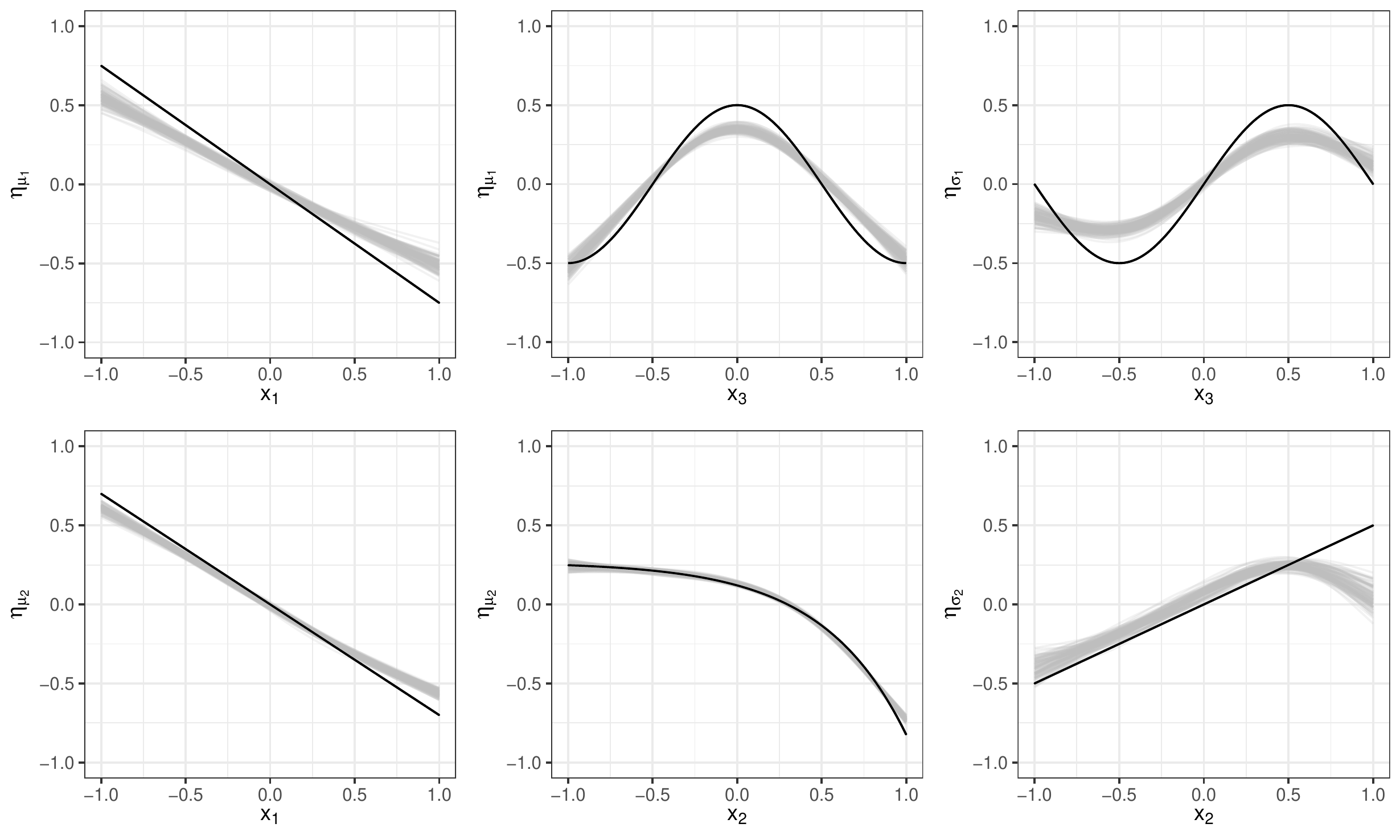}
\caption{Effect estimates on the marginal models for the Gauss copula in the high-dimensional setting. The first and the second row of plots show the informative covariate effects on the parameters of the first and the second marginal distribution (i.e.~Log-Normal and Log-Logistic distribution), respectively. In each plot the gray lines represent the estimates of the 100 runs while the black line displays the original effect.}
\label{fig:HighDimGauss}
\end{figure}

\clearpage

\subsubsection{Marginal estimates non-informative for Gauss, Clayton and Gumbel}\label{appen:HighDimMargEstNonInf}

\begin{figure}[H]
\includegraphics[width=0.95\textwidth]{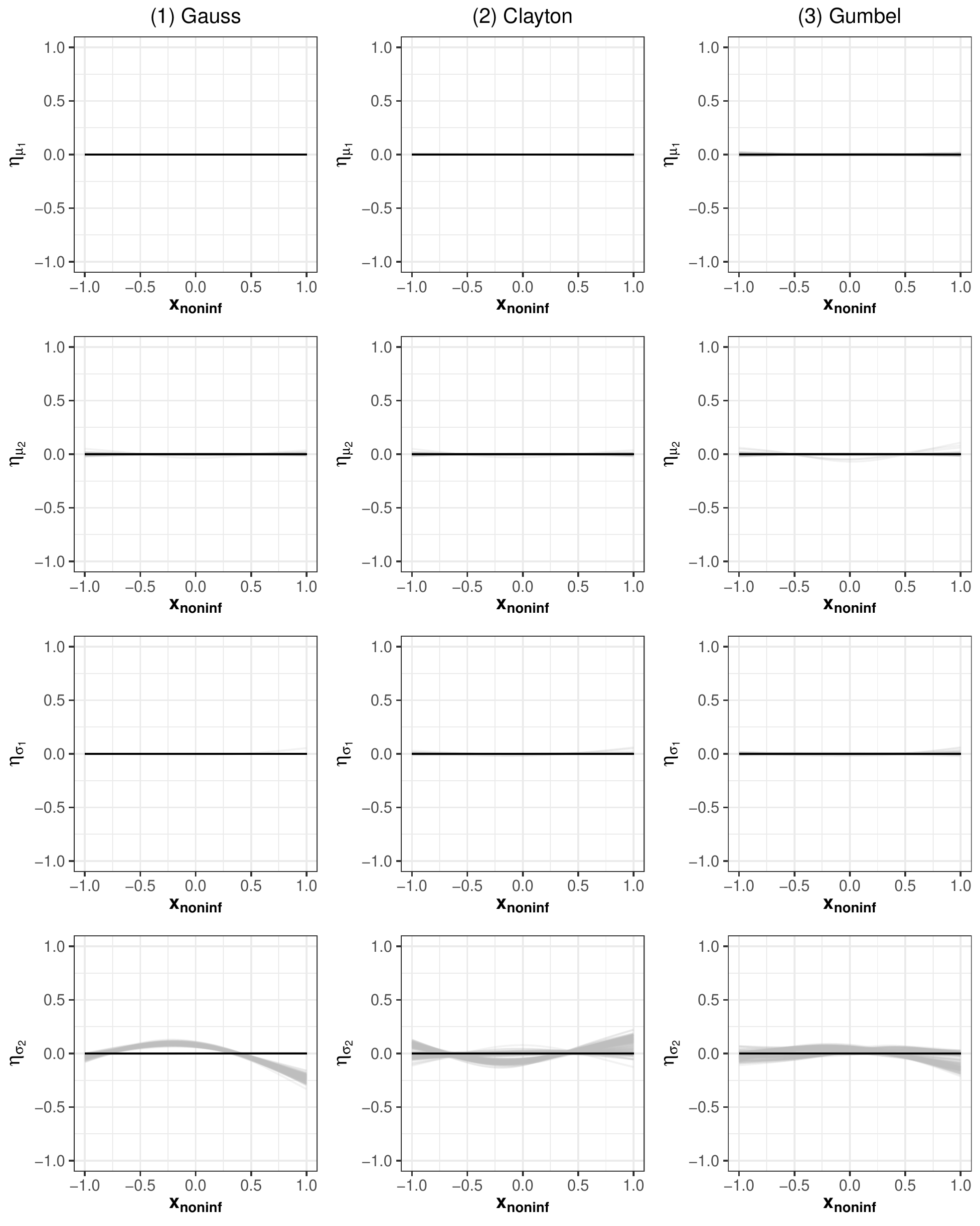}
\caption{Effect estimates of the non-informative covariates on the marginal distribution parameters for the Gaussian, Clayton and Gumbel copula in the high-dimensional setting. In each plot the gray lines represent the estimates of the 100 runs while the black line displays the original effect}
\label{fig:HighDimNonInf}
\end{figure}